\documentclass[useAMS,usenatbib,floats]{mn2e}
\usepackage[british]{babel}
\usepackage{graphicx}
\usepackage{subfig}
\usepackage{txfonts}
\usepackage{color}
\usepackage{mwe}
\usepackage{float}
\usepackage{caption}
\usepackage{tabularx}
\usepackage{blindtext}
\usepackage{varwidth}
\let\tilde=\widetilde
\title[Free-form modelling of galaxy clusters]{ Free-form modelling of
  galaxy clusters: a Bayesian and data-driven approach}
\label{firstpage}
\author[Olamaie et~al.]{Malak Olamaie,$^{1,2}$\thanks
{Email:m.olamaie@imperial.ac.uk} Michael P. Hobson$^{2}$,  
Farhan Feroz$^{2}$, Keith J. B. Grainge$^{2,4}$, \newauthor
Anthony Lasenby$^{2,3}$, Yvette C. Perrott$^{2}$, Clare Rumsey$^{2}$ and 
Richard D. E. Saunders$^{2,3}$\\
$^{1}$Imperial Centre for Inference and Cosmology (ICIC), Imperial College, Prince Concort Road, 
 London, SW7 2AZ\\
$^{2}$Astrophysics Group, Battcock Centre for Experimental Astrophysics,
 Cavendish Laboratory, 19 J. J. Thomson Avenue, Cambridge, CB3 0HE\\
$^{3}$Kavli Institute for Cosmology Cambridge, Madingley
      Road, Cambridge, CB3 0HA\\
 $^{4}$Jodrell Bank Centre for Astrophysics, School of Physics and Astronomy,
      The University of Manchester, M13 9PL}

\begin{document}

\date{Accepted ---; Received ---}

\pagerange{\pageref{firstpage}--\pageref{lastpage}}

\pubyear{2017}

\maketitle
\begin{abstract}
A new method is presented for modelling the physical properties of
galaxy clusters. Our technique moves away from the traditional
approach of assuming specific parameterised functional forms for the
variation of physical quantities within the cluster, and instead allows
for a `free-form' reconstruction, but one for which the level of
complexity is determined automatically by the observational data and
may depend on position within the cluster.  This is achieved by
representing each independent cluster property as some interpolating
or approximating function that is specified by a set of control
points, or `nodes', for which the number of nodes, together with their
positions and amplitudes, are allowed to vary and are inferred in a
Bayesian manner from the data.  We illustrate our nodal approach in
the case of a spherical cluster by modelling the electron pressure
profile $P_{\rm e}(r)$ in analyses both of simulated
Sunyaev--Zel'dovich (SZ) data from the Arcminute MicroKelvin Imager
(AMI) and of real AMI observations of the cluster MACS$\,$J0744+3927 in the
CLASH sample. We demonstrate that one may indeed determine the
complexity supported by the data in the reconstructed $P_{\rm e}(r)$,
and that one may constrain two very important quantities in such an
analysis: the cluster total volume integrated Comptonisation parameter
($Y_{\rm tot}$) and the extent of the gas distribution in the cluster
($r_{\rm max}$). The approach is also well-suited to detecting
clusters in blind SZ surveys.
\end{abstract}
\begin{keywords}
galaxies: cluster-- cosmology: observations -- methods: data analysis
\end{keywords}
%
\section{Introduction}
\label{sec:intro}
Determining the properties of clusters of galaxies, such as their
total and baryonic mass, has the potential to provide an independent
tool for constraining the parameters of the $\Lambda \rm{CDM}$ model,
since cluster population properties are sensitive to several
cosmological parameters, most notably $\sigma_8$ (see e.g.
\citealt{2013JCAP...10..060S}, \citealt{2016A&A...594A..24P}, and
\citealt{2016ApJ...832...95D}).  The challenge lies, however, in
obtaining a robust estimate of the cluster masses as these are not
directly observable. The mass distribution within a cluster is usually
measured using a variety of observational methods, including X-ray
(see e.g.  \citealt{2009ApJ...692.1060V}, \citealt{2013MNRAS.435.1265E},  
\citealt{2014MNRAS.440.2077M}
 and \citealt{2015MNRAS.446.1799O}), the
Sunyaev--Zel'dovich (SZ) effect (see e.g.  \citealt{2013A&A...550A.128P},
\citealt{2013MNRAS.431..900S}, \citealt{2013JCAP...07..008H},
\citealt{2015A&A...580A..95P}, \citealt{2016MNRAS.460..569R}, and 
\citealt{2017MNRAS.464.2378R}), and gravitational lensing (see e.g.
\citealt{2010ApJ...708..645R}, \citealt{2013SSRv..177...75H},
\citealt{2014MNRAS.438...78R} and \citealt{2016JCAP...08..013B}).

Each of these approaches relies on developing some method for
determining the cluster's mass (distribution) from its observable
properties, namely its distributions of X-ray surface brightness, SZ Comptonisation
parameter, and weak-lensing shear distribution.  This is
usually achieved by modelling the physical properties through the
cluster in terms of some specific parameterised functional
forms. Typical examples include assuming an NFW profile
\citep{1996ApJ...462..563N, 1997ApJ...490..493N} for the dark matter
density distribution, a $\beta$-model \citep{1976A&A....49..137C, 1978A&A....70..677C} 
for the gas density,
or a generalised NFW (GNFW) profile \citep{2007ApJ...668....1N} for
the gas pressure.  The cluster mass (distribution) is then usually
calculated under the standard assumption of a spherical cluster model
obeying hydrostatic equilibrium (HSE) and/or some scaling
relationships.

Even for physically-based cluster models (see
e.g. \citealt{2012MNRAS.423.1534O, 2013MNRAS.430.1344O}), there still
remains considerable uncertainty regarding the appropriate
form one should assume for the radial variation of cluster properties,
and this can lead to different cluster mass estimates (see e.g.
\citealt{2012MNRAS.421.1136A}, \citealt{2013SSRv..177..247G}, and
\citealt{2015MNRAS.453.3107K}, \citealt{2016MNRAS.461.3222D}). This
may result from either adopting inappropriate functional forms, often
by extrapolating their use to cluster masses and/or redshifts that are
not well sampled by observations or simulations, or by fitting models
that depend on parameters to which the data are insensitive.

In this paper, we therefore move away from the traditional approach of
assuming specific parameterised forms for the variation of cluster
properties, such as the pressure, density and/or temperature
distribution, and instead allow for a `free-form' reconstruction, but
one for which the level of complexity is determined automatically by
the observational data and may depend on position within the cluster.
This is achieved by representing each independent cluster property as
some interpolating or approximating function (for example, a piecewise
linear interpolation or a spline) that is specified by a set of
control points, or `nodes'. The positions and amplitudes of these
nodes and, most importantly, the number of nodes used are allowed to
vary and constitute the set of parameters to be inferred from the data
in a Bayesian manner.

We note that we have already successfully applied such a Bayesian
nodal modelling approach to a number of cosmological analysis
problems, including the reconstruction of the primordial
anisotropy power spectrum and the variation of the dark energy
equation-of-state parameter with redshift (see e.g.
\citealt{2012JCAP...09..020V}a,b and 
\citealt{2016MNRAS.455.2461H, 2017MNRAS.466..369H} ).

The structure of this paper is as follows. In Section~\ref{sec:Bayes},
we briefly summarise Bayesian inference, in particular parameter
estimation and model selection. Our nodal approach to modelling galaxy
clusters is presented in Section~\ref{sec:model}, and in
Section~\ref{sec:szapp} we apply it to the particular case of
modelling the pressure profile $P_{\rm e}(r)$ in a spherical cluster
observed via its SZ effect.  Section~\ref{sec:analysis} outlines our
Bayesian methodology for inferring the cluster parameters from
interferometric SZ observations, and summarises our simulated cluster
observations and real observations of the cluster MACS$\,$J0744+3927 using
AMI.  In Section~\ref{sec:Results},
we present the results of our Bayesian nodal analysis of these
simulations and real observations, and we conclude in
Section~\ref{sec:concs}.
\section{Bayesian inference}\label{sec:Bayes}
For the analysis of some data $\mathbf D$ in the context of a model (or
hypothesis) $\mathcal{M}$ that depends on some set of parameters
$\mathbf\Theta$, Bayes' theorem states that
\begin{equation}\label{eq:bayeseq}
  \Pr({\mathbf\Theta}|{\mathbf D},\mathcal{M}) =
    \frac{\Pr({\mathbf D}|{\mathbf\Theta}, \mathcal{M})
   \Pr({\mathbf\Theta}|\mathcal{M})}{\Pr({\mathbf D}|\mathcal{M})} \equiv \frac{\mathcal{L}(\mathbf
      \Theta)\pi(\mathbf \Theta)}{\mathcal{Z}}\,,
\end{equation}
which is usually interpreted as the prior probability distribution
$\Pr({\mathbf\Theta}|\mathcal{M})\equiv \pi(\mathbf\Theta)$ of the
parameters being updated by the likelihood $\Pr({\mathbf
  D}|{\mathbf\Theta},\mathcal{M})\equiv \mathcal{L}(\mathbf \Theta)$
of obtaining the observed data given some set of parameter values to
yield the posterior probability distribution $\Pr({\mathbf
  \Theta}|{\mathbf D},\mathcal{M})$ of the parameters, which is
normalised by the Bayesian evidence $\Pr({\mathbf
  D}|\mathcal{M})\equiv \mathcal{Z}$ (which does not depend on the
parameters $\mathbf \Theta$). The joint (unnormalised) posterior
provides the complete inference in Bayesian parameter estimation, and
can be subsequently marginalised over each parameter to obtain
individual parameter constraints.

Similarly, in Bayesian model selection, one can calculate the
probability of a model given the data as
\begin{equation}\label{eq:bayeseq2}
  \Pr(\mathcal{M}|{\mathbf D}) =
    \frac{\Pr({\mathbf D}|\mathcal{M})
   \Pr(\mathcal{M})}{\Pr({\mathbf D})} 
\equiv \frac{\mathcal{Z}\pi_\mathcal{M}}{\Pr({\mathbf D})}\,,
\end{equation}
Taking the ratio of the probabilities of two models signifies our
degree of belief in one model over another, and is given by
the posterior odds ratio (POR)
\begin{equation}\label{eq:Rval}
\mathcal{P}_{ij}\equiv\frac{\Pr(\mathcal{M}_j|\mathbf
  D)}{\Pr(\mathcal{M}_i|\mathbf D)} = 
\frac{Z_j\pi_{\mathcal{M}_j}}{Z_i\pi_{\mathcal{M}_i}} =
\mathcal{B}_{ij}
\frac{\pi_{\mathcal{M}_j}}{\pi_{\mathcal{M}_i}}\,,
\end{equation}
where $\mathcal{B}_{ij}\equiv \mathcal{Z}_j/\mathcal{Z}_i$ is the Bayes factor
\citep{1961xrec.book.....J}. Clearly, if
$\pi_{\mathcal{M}_j}=\pi_{\mathcal{M}_i}$, so that the two models are
considered {\em a priori} equally probable, then $\mathcal{P}_{ij} =
\mathcal{B}_{ij}$.  Table~\ref{tab:bayesfactor} lists a modern version
of Jeffreys' criteria, which are used to give meaning to this
quantification \citep{1995Journal of the American Statistical
  Association...90...430 , 2013 IEEE 13th International
  Conference...10.1109/ICDMW.2013.21}.
\begin{table}
\caption{Jeffreys' scale for interpreting PORs (or Bayes factors).
As $\ln\mathcal{P}_{ij} = -\ln\mathcal{P}_{ji}$ negative values imply
reversed model favouring.\label{tab:bayesfactor}}
\begin{center}
\begin{tabular}{@{}ll@{} }\hline
$\ln\mathcal{P}_{ij}$ & Favouring of $\mathcal{M}_j$ over $\mathcal{M}_i$
\\\hline
$0.0 \leq \ln\mathcal{P}_{ij} < 1.0$ & None \\
$1.0 \leq \ln\mathcal{P}_{ij} < 3.0$ & Slight \\
$3.0 \leq \ln\mathcal{P}_{ij} < 5.0$ & Significant \\
$5.0 \leq \ln\mathcal{P}_{ij}$ & Decisive\\\hline
\end{tabular}
\end{center}
\end{table}

Thus, for Bayesian model selection, in contrast to parameter
estimation, the evidence takes a central role. Typically, the
evidence for each model is calculated separately and their ratios
evaluated. From (\ref{eq:bayeseq}), the evidence is the normalisation
constant for the posterior, which is given by
\begin{equation}\label{eq:evidence}
  \mathcal{Z}=\int\mathcal{L}(\mathbf \Theta)\pi(\mathbf \Theta)
  \, \mathrm{d}^D\mathbf\Theta\,,
\end{equation}
where $D$ is the dimensionality of the parameter space. As the average
of the likelihood over the prior, the evidence embodies the notion of
Occam's razor (see e.g, \citealt{1986pps.book.....J} and
\citealt{2005xrec.book.....S}): a simple theory with a compact parameter
space will have a larger evidence than a more complicated one, unless
the latter is significantly better at explaining the data.

The evaluation of the multidimensional integral (\ref{eq:evidence})
over the whole parameter space is a challenging numerical task. We
perform this calculation here using the nested sampling algorithm
\textsc{Multinest} \citep{2008MNRAS.384..449F,2009MNRAS.398.1601F,
  2013arXiv1306.2144F}.  This Monte-Carlo method is targeted at the
efficient calculation of the evidence, but as a by-product also
produces posterior inferences for parameter estimation; it is also
very efficient at exploring posteriors that contain multiple modes
and/or large (curving) degeneracies.

We note that elsewhere \citep{2016MNRAS.455.2461H,2017MNRAS.466..369H}, 
we have presented an alternative
method for performing Bayesian model selection, without explicitly
computing evidences, which uses a combined likelihood and introduces
an integer model selection parameter $n$.  If the total number of
models under consideration is specified {\em a priori}, the full joint
parameter space of the models is of fixed dimensionality and can be
explored using standard sampling methods, without the need for
trans-dimensional techniques, although the posterior is usually highly
multimodal and hence nested sampling is again an obvious choice. Bayes
factors, or more generally posterior odds ratios, may then be read off
directly from the posterior on $n$, which is obtained by
straightforward marginalization. To keep our discussion simple,
however, we will not use this method here, but plan to apply it to
nodal modelling of galaxy clusters in a forthcoming publication.

In closing this section, it should be mentioned that, in general, a
gain in information via a Bayesian analysis may be achieved in several
ways: it can occur because of a tightening of the parameter
constraints, a shift in position of the peak(s) of the distribution
from prior to posterior, or an increase in the evidence (see
e.g. \citealt{2008JHEP...12..024T}, \citealt{2014PhRvD..90b3533S} and
\citealt{2016PhRvD..93j3507S}).

\section{Nodal model for a galaxy cluster}
\label{sec:model}

As described briefly earlier, in our nodal approach to free-form
modelling of a galaxy cluster, each independent physical property of
the cluster is represented by some interpolating or approximating
function that is determined by a set of control points, or nodes.  In
principle, these functions can be fully three-dimensional to allow
modelling of arbitrary structure in each property of interest in the
cluster. To illustrate the method simply, however, we will consider
here the special case of a spherical cluster, such that each property
is a function only of radius $r$ from the cluster centre. Moreover, we
will specialise still further to the case where one constructs a nodal
model of just a single property of interest, described by some
function $f(r)$. It is a straightforward matter to extend the
following analysis to multiple properties of interest.

\begin{figure}
\begin{center}
\includegraphics[width=0.9\linewidth]{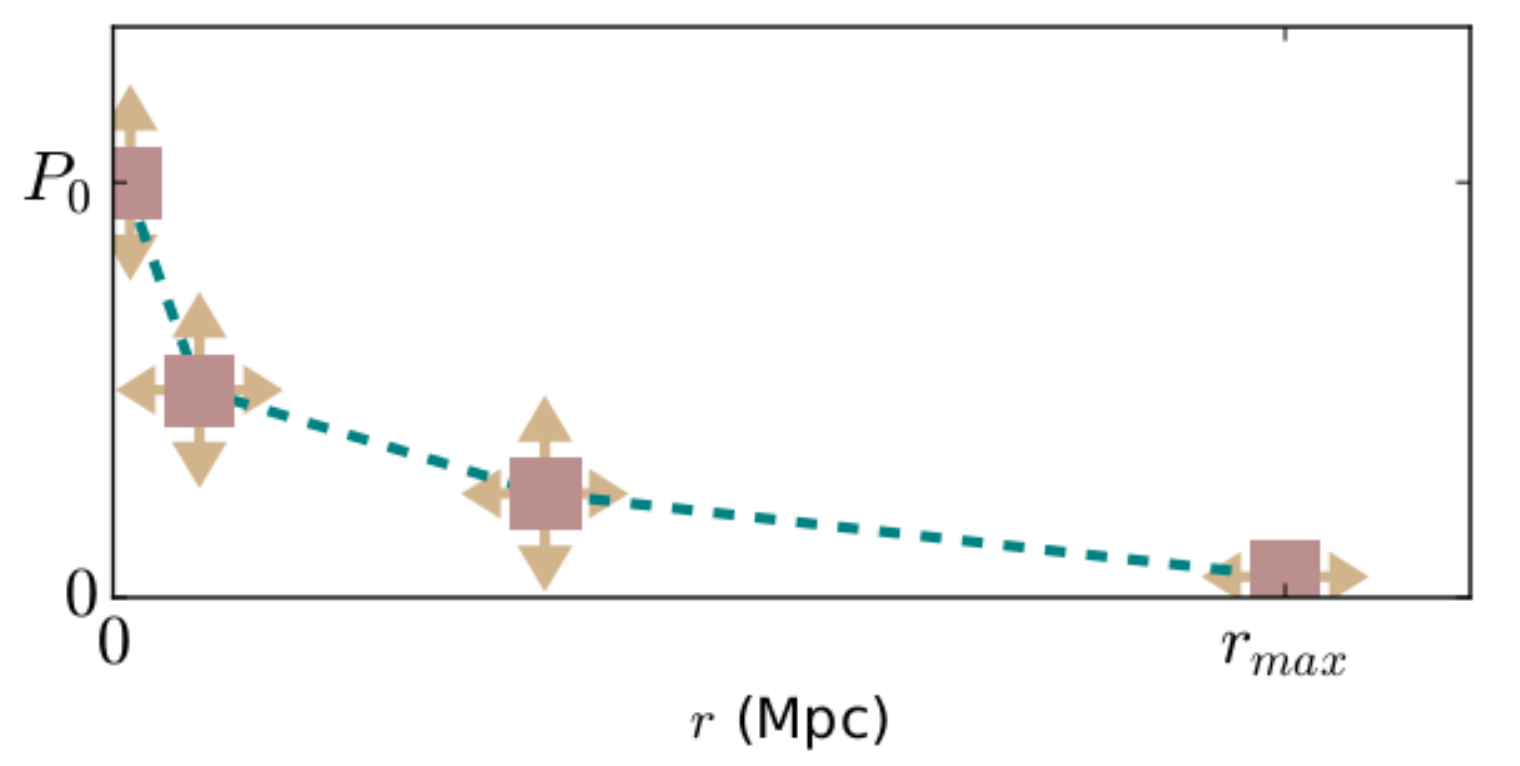} 
\caption{Linearly-interpolated nodal representation for $N=4$ nodes of 
the gas pressure profile $P_{\rm e}(r)$ in a galaxy cluster described in 
units of  $\rm{MeVm}^{-3}$  ($1\rm{MeVm}^{-3}=1.602\times10^{-13}\rm{Nm}^{-2}$).
  \label{fig:Nodalmodel}}
\end{center}
\end{figure}

The basic idea is to represent $f(r)$ not with some standard
parameterised functional form, as in most current cluster analyses,
but in terms of a number $N$ of nodes in $(r,f)$-space, together with
their corresponding positions $r_n$ and amplitudes $f_n$
$(n=0,1,2,\ldots,N-1)$ (thus each node can `move' both horizontally and
vertically-- see figure~\ref{fig:Nodalmodel}; although we will in fact restrict the movement of the
`end' nodes, as described below).  The $N$ nodes act as control points
for the continuous function $f(r)$.  In this way, one obtains a
continuous free-form reconstruction of the profile $f(r)$, for which the
complexity is regularised by the data under analysis.

One is free to choose from a wide range of possible interpolation or
approximation methods, such as polynomials, rational functions,
splines, B\'ezier curves or even Gaussian processes. It should be
noted, however, that some forms of smooth interpolating or
approximating function, such as cubic splines, have continuity
requirements on the function and its derivatives which can
significantly reduce the ability of the resulting $f(r)$ to reproduce
abrupt features in the cluster profile \citep{2012JCAP...06..006V}. In
principle, the nature of the interpolating or approximating function
could be determined by performing a straightforward Bayesian model
selection between the various options, although we will not consider
that further here. Instead, for illustration, we choose here the
simplest approach of linearly interpolating between the nodes; since
we are performing an interpolation, rather than an approximation, one has
$f_n = f(r_n)$.

As mentioned above, given the nature of the one-dimensional function
$f(r)$ that we wish to model in the case of a galaxy cluster, we
typically restrict the movement of the first and last nodes (or end
nodes) as follows. The first node has a fixed position, at the origin,
such that $r_0 = 0$ always. Consequently, its corresponding amplitude
parameter $f_0$ represents the central value $f(0)$, which is often of
interest in galaxy cluster analyses. In contrast, the last node has a
fixed amplitude of zero, such that $f_{N-1}=0$. Thus, its
corresponding position parameter $r_{N-1}$ can sometimes be
interpreted as an extent of the cluster $r_{\rm max}$, which is again
often of interest (although this interpretation does depend on the
nature of the quantity $f(r)$ being modelled). In our demonstration of
the method presented in Section~\ref{sec:szapp}, $f(r)$ represents the
electron pressure profile $P_{\rm e}(r)$ of the cluster and so
$r_{N-1}$ represents the extent of the cluster
gas. Figure~\ref{fig:Nodalmodel} illustrates our linearly-interpolated
nodal model of $P_{\rm e}(r)$ for $N=4$ nodes.

In the context of Bayesian inference, we denote the model consisting
of $N$ nodes by $\mathcal{M}_N$, which has the parameters
${\mathbf\Theta} = \{r_n,f_n\}$ for $n=0,1,2,\ldots,N-1$ (with delta
function priors on $r_0$ and $P_{N-1}$), plus any
further `global' parameters, such as the cluster position on the sky
or parameters describing any contaminating signals or noise in the
data.  The priors on the parameters may be chosen to accommodate
whatever information is available {\em a priori}. In general, however,
we typically also impose a `sorting condition' on the node positions,
such that $r_n < r_{n+1}$. This can be done straightforwardly, without
the need to reject any samples. Indeed, if each node position is
considered to be drawn from a uniform distribution in some given
range, as is usually the case, the corresponding (non-separable) joint
prior $\pi(r_0,r_1,r_2,\ldots,r_{N-1})$ on the node positions, which
incorporates the sorting condition $r_n < r_{n+1}$, has an analytic
form in terms of the beta distribution (\citealt{2015MNRAS.453.4384H}).  It
is also worth mentioning that, although we will assume here that each
model $\mathcal{M}_N$ is equally likely {\em a priori}, such that
$\pi_{\mathcal{M}_i} = \pi_{\mathcal{M}_j}$ (and hence PORs are
equivalent to Bayes factors), this is not necessary. One may view this
assumption as imposing a uniform prior (within some range) on the
number $N$ of nodes, but one could equally well impose, for example, a
Poisson prior on $N$ with some given mean, from which one can read off
the corresponding values of $\pi_{\mathcal{M}_N}$.

In the straightforward model selection approach used here, one begins
by analysing the $N=2$ model, using \textsc{Multinest} to calculate
the evidence $\mathcal{Z}_2$ and obtain samples from the posterior
$\Pr({\mathbf\Theta}|{\mathbf D},\mathcal{M}_2)$. This process is then
repeated separately for $N=3,4,\ldots$ until the evidence
$\mathcal{Z}_N$ has decreased well below that of the most favoured
model. One may then proceed either by conditioning on the most
favoured model $\mathcal{M}_{\hat{N}}$ and simply infer parameters and
the corresponding constraints on $f(r)$ from the posterior
$\Pr({\mathbf\Theta}|{\mathbf D},\mathcal{M}_{\hat{N}})$ or perform a
`multi-model' inference 
(see e.g. \citealt{2013SADM....6....3P}, \citealt{2016MNRAS.455.2461H, 2017MNRAS.466..369H}, 
and \citealt{2016A&A...594A..20P}) 
in which the
constraints on $f(r)$ are determined by averaging over all the models
$\mathcal{M}_N$ considered, weighted by their PORs. In the interests
of simplicity we will adopt the former approach here.

\section{Application to SZ observations}
\label{sec:szapp}

Although our nodal approach may be used to model observations of
galaxy clusters in any waveband, we illustrate the method here by
applying it to SZ observations  (see e.g.
\citealt{1970CoASP...2...66S}, \citealt{1999PhR...310...97B}, and
\citealt{2002ARA&A..40..643C}). 
Since the SZ surface brightness is
proportional to the line-of-sight integral of the electron pressure,
as we show below, it is natural to use the nodal approach to model the
radial electron pressure profile $P_{\rm e}(r)$ of the ionised gas within the cluster.

The observed SZ surface brightness $\delta I_\nu$ in the direction of
an electron reservoir is given by
\begin{equation}\label{deltaI}
\delta I_\nu=T_{\rm CMB}yf(\nu)\frac{\partial B_\nu}{\partial T}\Big\vert_{T=T_
{\rm CMB}}\,.
\end{equation}
Here $B_\nu$ is the blackbody spectrum, $T_{\rm CMB}=2.73 $~K (Fixsen
et~al.  1996) is the temperature of the CMB radiation,
$f(\nu)=\left(x\coth(x/2)-4\right)(1 + \delta (x , T_{\rm e}))$ is the
frequency dependence of thermal SZ signal, $x=\frac{h_{\rm
    p}\nu}{k_{\rm B}T_{\rm CMB}}$, $h_{\rm p}$ is Planck's constant,
$\nu$ is the frequency and $\rm{k_{\rm B}}$ is Boltzmann's
constant. The function $\delta (x , T_{\rm e})$ takes into account the
relativistic corrections in the study of the thermal SZ effect which
is due to the presence of thermal weakly relativistic electrons in the
ICM and is derived by solving the Kompaneets equation up to the higher
orders (\citealt{1995ARA&A..33..541R}, \citealt{1998ApJ...502....7I},
\citealt{1998ApJ...508...17N}, \citealt{1998A&A...336...44P},
\citealt{1998ApJ...499....1C}). It should be noted that at 15 GHz (AMI
observing frequency) $x= 0.3$ and therefore the relativistic
correction, as shown by \cite{1995ARA&A..33..541R}, is negligible for
$k_{\rm B}T_{\rm e} \leq 15\, \rm {keV}$. The dimensionless parameter
$y$, known as the Comptonization parameter, is the integral of the
number of collisions multiplied by the mean fractional energy change
of photons per collision, along the line of sight
\begin{equation}\label{eq:ypar}
 y = \frac{\sigma_{T}}{m_{\rm e}c^2} \int_{-\infty}^{+\infty}{P_{\rm e}(r)\,{\rm 
d}l}\,,
\end{equation}
where $P_{\rm e}(r)$ is the electron pressure at radius $r$
respectively, $\sigma_{\rm T}$ is Thomson scattering cross-section,
$m_{\rm e}$ is the electron rest mass, $c$ is the speed of light and $dl$
is the line element along the line of sight.  It should be noted that
(\ref {eq:ypar}) assumes an ideal gas equation of state.

The integral ($Y_{\rm SZ}$) of the Comptonisation $y$ parameter over the solid 
angle $\Omega$ subtended by the cluster is proportional to the 
volume integral of the gas pressure. It is thus a good estimate for the 
total thermal energy content of the cluster and hence its mass (see e.g.
Bartlett \& Silk 1994). The $Y_{\rm SZ}$ parameter in cylindrical and 
spherical geometries, respectively, may be described as
\begin{eqnarray}\label{eq:Ycylsph}
Y_{\rm cyl}(R)&=& \frac{\sigma_{T}}{m_{\rm e}c^2}\int_{-\infty}^{+\infty}
{\rm d}l\,\int_{0}^{R}{P_{\rm e}(r)2\pi s \, {\rm d}s}\,, \\
Y_{\rm sph}(r)&=& \frac{\sigma_{\rm T}}{m_{\rm e}c^2}\int_{0}^{r}{P_{\rm 
e}(r')4\pi r'^{2}{\rm d}r'},  
\end{eqnarray}
where $R$ is the projected radius of the cluster on the sky. Thus, by
using our nodal approach to model $P_{\rm e}(r)$, one may constrain
$Y_{\rm SZ}$ in either geometry.

\section{Analysis of interferometric SZ data}
\label{sec:analysis}

In order to verify that our proposed model, with its corresponding
assumptions, can describe profiles of cluster physical properties
accurately, we carry out a Bayesian analysis of simulated SZ
observations using AMI \citep{2008MNRAS.391.1545Z} 
of a set of three clusters, as well as real
AMI SZ observations of the cluster MACS$\,$J0744+3927 \citep{2016MNRAS.460..569R}.

\subsection{Bayesian methodology}

An interferometer like AMI operating at a frequency $\nu$ measures
samples from the complex visibility plane $\tilde{I}_\nu({\bf
  u})$. These are given by a weighted Fourier transform of the surface
brightness $I_\nu({\bf x})$, namely
\begin{equation}\label{eq:Ivis}
  \tilde{I}_\nu({\bf u})=\int{A_\nu({\bf x})I_\nu({\bf x})\exp(2\pi i{\bf u\cdot x})\,
{\rm d}{\bf x}}\,,
\end{equation}
where ${\bf x}$ is the position on the sky relative to the phase centre, $A_\nu({\bf x})$
is the (power) primary beam of the antennas at observing frequency $\nu$
(normalised to unity at its peak) and ${\bf u}$ is the baseline vector in units
of wavelength. 

Details of our Bayesian methodology for modelling interferometric SZ
data, primordial CMB anisotropies, and resolved and unresolved radio
point-source are given in
\cite{2002MNRAS.334..569H,2008MNRAS.384..449F,2009MNRAS.398.2049F,
2011MNRAS.415.2708A} and \cite{2012MNRAS.419.2921A}.

In short, the measured interferometer visibilities in our model are
assumed to have the form
\begin{equation}\label{eq:Vvis}
  V_\nu({\bf u})=\tilde{I}_\nu({\bf u}) + N_\nu({\bf u})\,,
\end{equation}
where the signal part, $\tilde{I}_\nu({\bf u})$, contains the
contributions from the SZ cluster and identified radio point sources,
whereas the generalised noise part, $N_\nu({\bf u})$, contains
contributions from the background of unsubtracted radio point sources,
primary CMB anisotropies and instrumental noise. The last two
contributions to the generalised noise are well described by Gaussian
processes, whereas the background of unsubtracted radio sources is
strictly a Poisson process. Nonetheless, in the limit of a large
number of faint unsubtracted sources, this contribution can also be
well approximated as Gaussian (Feroz et al. 2009). Thus, the
generalised noise is assumed to be Gaussian distributed, so that the
likelihood function has the form
\begin{equation}\label{eq:like}
  \mathcal{L}(\mbox{\boldmath $\Theta$}) \propto
  \exp\left(-{\textstyle\frac{1}{2}}\chi^2 \right)\,,
\end{equation}
where $\chi^2$ is the standard statistic quantifying the misfit
between the observed data $\mbox{\boldmath $D$}$ and the predicted
data $\mbox{\boldmath $D$}^{\rm p}(\mbox{\boldmath$\Theta$})$:
\begin{equation}\label{eq:chisq}
  \chi^2=\sum_{\nu , \nu' }(\mbox{\boldmath $D$}_\nu -\mbox{\boldmath
    $D$}_\nu^{\rm p})^T
(\mbox{\boldmath $C$}_{\nu , \nu'})^{-1}(\mbox{\boldmath $D$}_{\nu'}-\mbox{\boldmath 
$D$}_{\nu'}^{\rm p})\,,
\end{equation}
in which $\mbox{\boldmath $C$}_{\nu , \nu'}$ is the generalised noise
matrix relating the frequency channels $\nu$ and $\nu'$. Under the
assumption that the three contributions to the generalized noise
discussed above are independent, this matrix is simply the sum of the
covariance matrices for each component. These matrices are described
in Feroz et al. (2009) and are assumed to be independent of the
parameters to be fitted in the analysis.  In particular, the primary
CMB anisotropies are assumed to be consistent with the concordance
cosmology.

In the Bayesian analysis, the point sources are modelled using
delta-function priors on position and Gaussian priors on flux and
spectral index, usually determined from higher-resolution observations
from the AMI Large Array, as discussed in Feroz et al. (2009).  We
focus here, however, on the inference of the cluster parameters. For
the model $\mathcal{M}_N$, having $N$ nodes, the cluster parameters
are 
\begin{equation}
\mbox{\boldmath $\Theta$}_{\rm c}\equiv (x, y, r_0 , P_0, \ldots,
r_{N-1},P_{N-1})\,,
\end{equation}
where $x$ and $y$ are the cluster projected
position on the sky.  We carry out the analysis for
$N=2,3,4,\ldots,8$.  We assume that the priors on these sampling
parameters are separable, apart from imposing the `sorting condition'
on the nodes positions, as discussed in Section~\ref{sec:model}, so that
\begin{equation}\label{eq:prior1}
 \pi(\mbox{\boldmath$\Theta$}_{\rm
   c})=\pi(x)\,\pi(y)\,\pi(r_0,r_1,\ldots,r_{N-1})\,\prod_{n=0}^{N-1}
 \pi(P_n)\,.
\end{equation}
The position of the first node is fixed to $r_0=0$, whereas the
position of the last node has a uniform prior in the range $0.5 \le
r_{N-1}/\mbox{Mpc} \le 2.5$. The positions of the intervening nodes
have uniform priors in the range $0 \le r_n/\mbox{Mpc} \le 1$. The
prior on the pressure $P_0$ of the first node is a truncated
exponential distribution with mean $\lambda=100$ MeVm$^{-3}$ in the
range $10 \le P_0/\rm{MeVm}^{-3} \le 5 \times 10^3$, whereas the
pressure $P_{N-1}$ of the last node is fixed to zero. The pressures
$P_n$ of the intervening nodes have uniform priors in the range $0 \le
P_n/\rm{MeVm}^{-3}\le 500$. Finally, we assume Gaussian priors on the
cluster position parameters $x$ and $y$, centred on the origin with a
standard deviation of 1 arcmin.  It is also worth mentioning that we
assume here that each model $\mathcal{M}_N$ is equally likely {\em a
  priori}, such that $\pi_{\mathcal{M}_i} = \pi_{\mathcal{M}_j}$ (and
hence PORs are equivalent to Bayes factors), which is equivalent to
imposing a uniform prior on $N$ in the range $2 \le N \le 8$.  We note
that the above priors are chosen to be consistent with the results of
$N$-body simulations and real cluster observations of galaxy clusters
(see e.g.\ \citealt{2012MNRAS.423.1534O},
\citealt{2013MNRAS.430.1344O} and references therein). In all
analyses, the redshift of the cluster is assumed known.

\subsection{Simulated AMI observations}

To generate simulated SZ skies and observe them with a model AMI
instrument, we use the methods outlined in \cite{2002MNRAS.334..569H},
\cite{2002MNRAS.333..318G}, \cite{2009MNRAS.398.2049F}, and
\cite{2012MNRAS.419.2921A}. In particular, we generate simulated
observations of three different model clusters, which we call
$\rm{SIM}_1$, $\rm{SIM}_2$ and $\rm{SIM}_3$. These differ in
the input pressure profile used to generate them, and all include
primordial CMB anisotropies, receiver thermal noise, and simulated residual 
points sources typical of AMI pointed observations of clusters.

\begin{figure}
\begin{center}
\includegraphics[height=6.5cm,clip=]{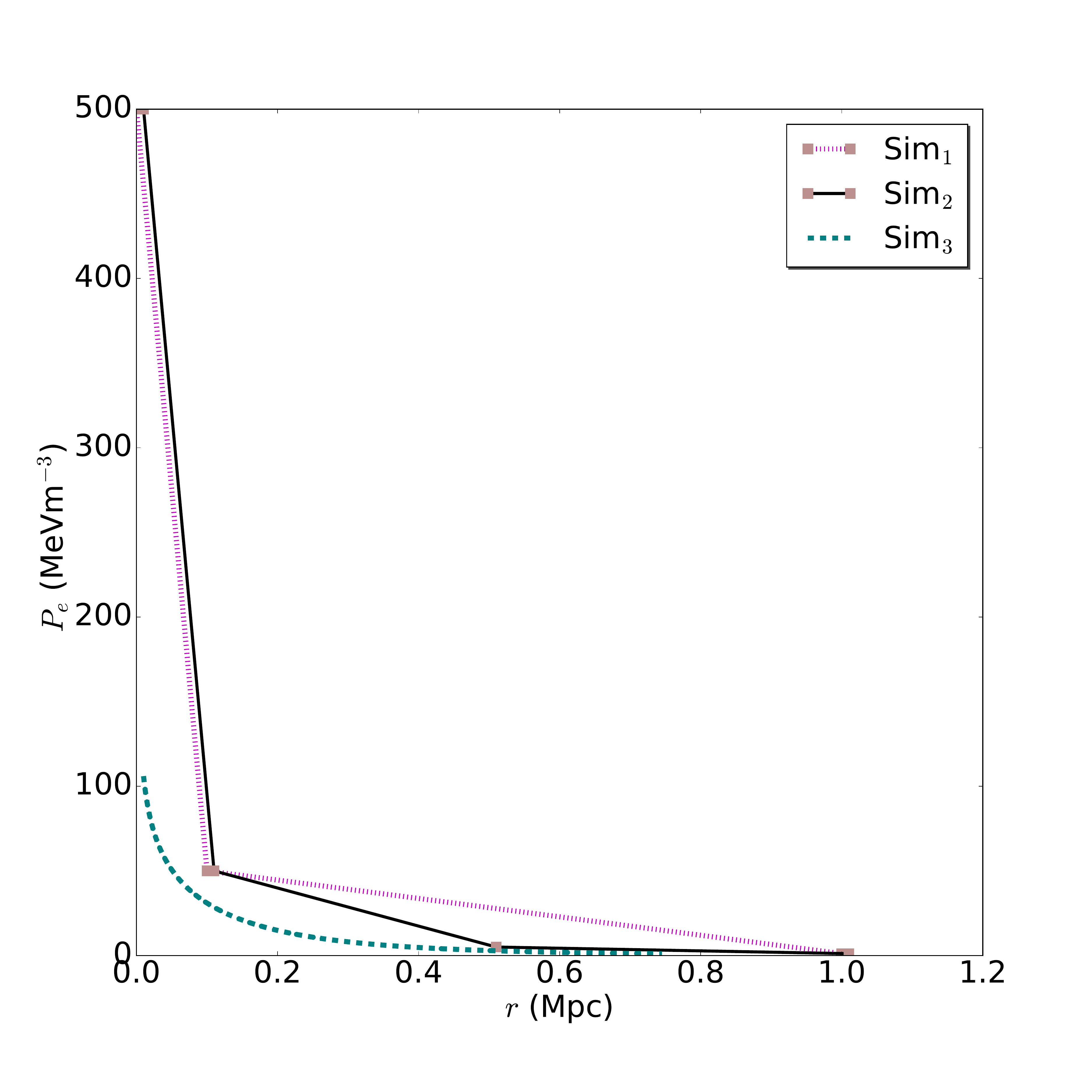} 
\caption{The input pressure profiles for the
  simulated clusters $\rm{SIM}_1$, $\rm{SIM}_2$ and $\rm{SIM}_3$.  The
  profiles for $\rm{SIM}_1$ and $\rm{SIM}_2$ were generated by linear
  interpolation between $N=3$ and $N=4$ nodes, respectively. The
  profile for $\rm{SIM}_3$ was generated using the model described in
  Olamaie et al. (2012, 2013), with $M_{\rm {tot}}(r_{\rm 200}) = 5
  \times 10^{14}\,\rm{M_\odot}$, $z=0.54$ and $f_{\rm
    {gas}}(r_{200})=0.13$.\label{fig:DMGNFWP}}
\end{center}
\end{figure}

For $\rm{SIM}_1$ and $\rm{SIM}_2$, we assume an input pressure profile
that matches our nodal model precisely, in order to investigate the
ability of our approach to recover the true parameter values in the
simplest case. In particular, the pressure profiles for $\rm{SIM}_1$
and $\rm{SIM}_2$ are generated by linear interpolation between $N=3$
and $N=4$ nodes, respectively, and are plotted in
figure~\ref{fig:DMGNFWP}. For both simulations, we assume the cluster
lies at a redshift of $z=0.5$.

For SIM$_3$, we use a more realistic cluster model to test the ability
of our nodal approach to recover a pressure profile that is not of the
form assumed in the analysis. In this case, the cluster is simulated
using the model described in \cite{2012MNRAS.423.1534O,
  2013MNRAS.430.1344O}, which assumes that the dark matter density
follows an NFW profile and the ICM plasma pressure is described by the
generalised NFW (GNFW) profile. The model also assumes that
hydrostatic equilibrium is satisfied and that the local gas fraction
is small throughout the cluster. This cluster model is fully specified
by just three parameters, for which we assume the values $M_{\rm
  {tot}}(r_{\rm 200}) = 5 \times 10^{14}\, \rm{M_\odot}$, $z=0.54$ and
$f_{\rm {gas}}(r_{200})=0.13$. The resulting pressure profile is shown
in figure~\ref{fig:DMGNFWP}, and is formally singular at the origin, 
decreasing sharply with radius in the central regions of the cluster.
\subsection{AMI observations of MACS$\,$J0744+3927}

We also analyse real AMI observations of MACS$\,$J0744+3927, one of the
clusters in the CLASH (Cluster Lensing And Supernova survey with Hubble)
sample \citep{2012ApJS..199...25P}.  MACS$\,$J0744+3927 is a rich cluster at
redshift z = 0.689 and has been studied through its X-ray emission,
strong lensing, weak lensing and SZ effect \citep{2007MNRAS.379..209S,
  2009A&A...496..343E, 2016ApJ...821..116U, 2016MNRAS.460..569R}. The SZ signal (decrement) on the AMI map appears
circular, (figure~7 in \citealt{2016MNRAS.460..569R}) in agreement
with the X-ray surface brightness from the Chandra archive data
(figure~6 in \citealt{2012ApJS..199...25P}).  Details of AMI pointed
observation towards the cluster, data reduction pipeline and mapping
are described in \cite{2016MNRAS.460..569R}. In particular, we note
that the Bayesian analysis includes 23 radio point sources in the AMI
field. We focus here, however, on the determination of the parameters
defining our nodal model of the cluster.

\section{Results and Discussion} 
\label{sec:Results}

We now present the results of our Bayesian nodal analysis applied to
the three simulated data sets $\rm{SIM}_1$, $\rm{SIM}_2$, and
$\rm{SIM}_3$, and real AMI observations of the cluster MACS$\,$J0744+3927.

\subsection{Simulation SIM$_1$}
\begin{figure}
\includegraphics[width=\linewidth]{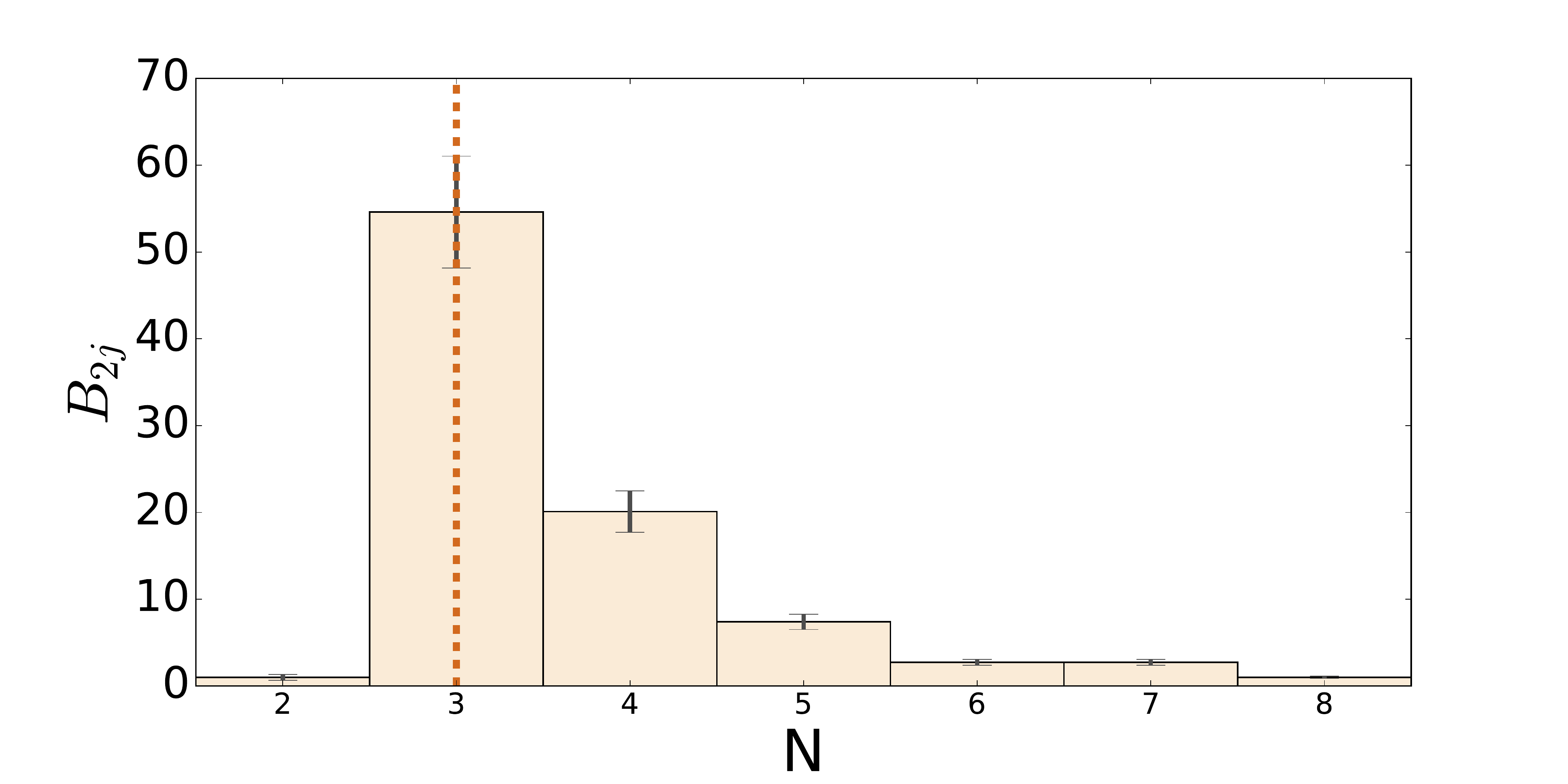}
\caption{Histogram of Bayes factors $\mathcal{B}_{2j}$ (i.e.\ relative
  to the $N=2$ model) as a function of the number of nodes $N$ in the
  reconstruction of the pressure profile $P_{\rm e}(r)$ for simulation
  SIM$_1$; this is equivalent to the marginalised posterior on
  $N$. The estimated errors on the Bayes factors are also shown. The
  vertical dotted line indicates the most favoured value of
  $N$.\label{fig:bfhist1}}
\end{figure}
\begin{figure*}
\includegraphics[width=0.48\linewidth]{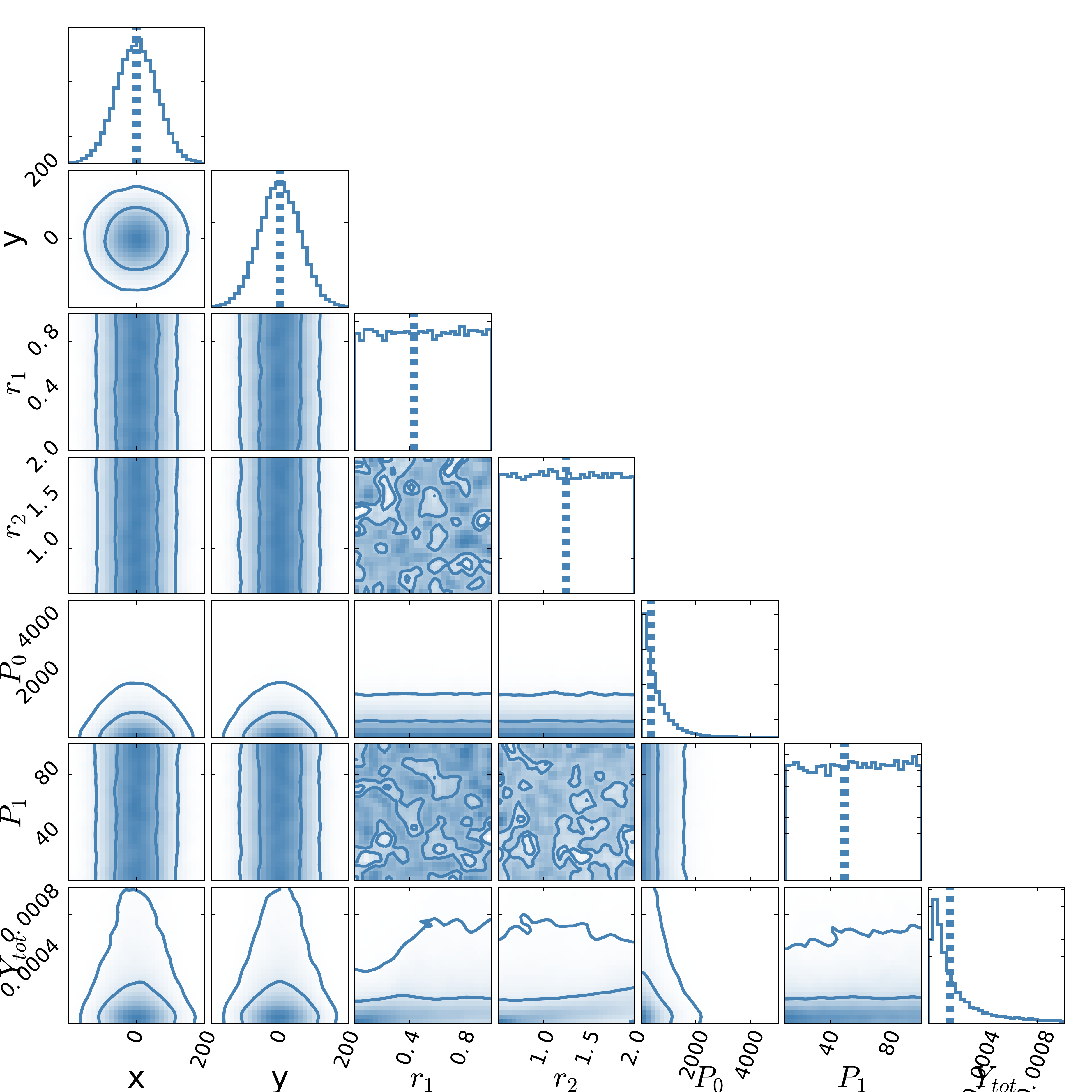}
\qquad\includegraphics[width=0.48\linewidth]{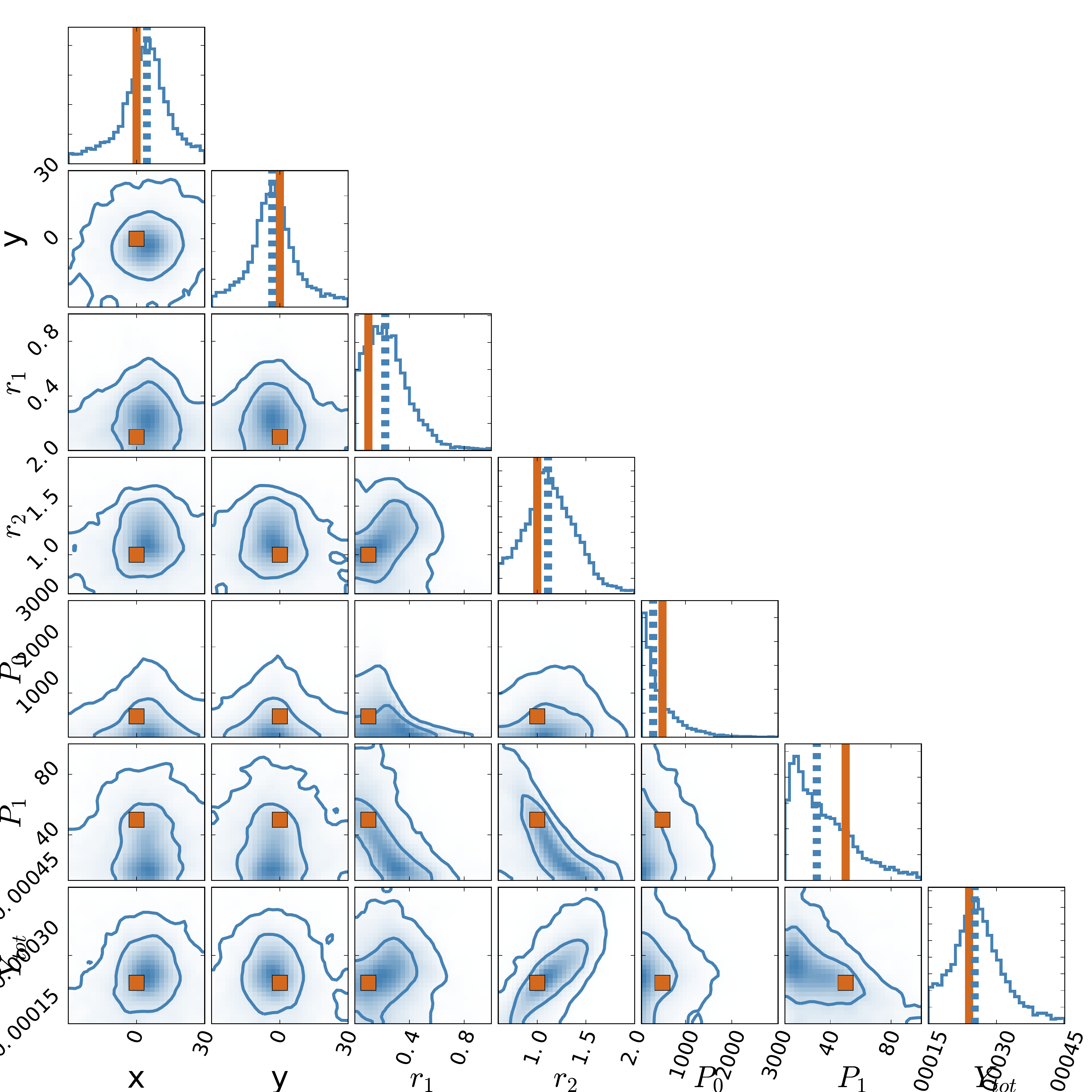}
\caption{1-D and 2-D marginal posterior distributions of the cluster
  sampling parameters ${\mathbf\Theta}_{\rm c}$ and the derived
  parameter $Y_{\rm tot}$, conditioned on $N=3$ nodes and obtained in the
  absence of any data (left) and from analysis
of simulation SIM$_1$ (right). $x$ and $y$ are in units of arcseconds, 
$r$s are in Mpc, $P$s are in $\rm{MeVm}^{-3}$ and $Y_{\rm tot}$ is in $\rm{Mpc}^2$. 
The contours on the 2-D distributions represent
  $68\%$ and $95\%$ Bayesian confidence intervals. The vertical dotted
  lines show the mean values of the 1-D
  distributions. 
The squares in the 2-D distributions and
the vertical solid lines in the 1-D distributions indicate the true
values of the parameters used in the simulation. Note that the
regions of the parameter space depicted in the right-hand plot
are typically much smaller than those in the left-hand one.\label{fig:post1}}
\end{figure*}

The results of our evidence-based Bayesian model selection analysis to
determine the number $N$ of nodes in the $P_{\rm e}(r)$ reconstruction
for $\rm{SIM}_1$ are given in figure~\ref{fig:bfhist1}. This shows the
histogram of Bayes factors $\mathcal{B}_{2j}$, i.e.\ relative to the
$N=2$ `straight line' model, as a function of the number of nodes $N$
in the reconstruction of the pressure profile $P_{\rm e}(r)$ for
simulation SIM$_1$. Given our prior choice
$\pi_{\mathcal{M}_i}=\pi_{\mathcal{M}_j}$ on the models, the Bayes
factors are equal to the PORs $\mathcal{P}_{2j}$, and so the plotted
histogram is equivalent to the marginalised posterior on $N$.
Recalling that the input pressure profile for SIM$_1$ is constructed
by linear interpolation between $N=3$ nodes, one sees that our
analysis has recovered the true value of $N$ as the most favoured.  In
particular, it is worth noting that the log Bayes factor
$\ln\mathcal{B}_{23}=4.0\pm 0.12$, indicating strong evidence for the
$N=3$ model over the $N=2$ (straight line) model, according to
Table~\ref{tab:bayesfactor}. The Bayes factors then gradually decline
for $N > 3$, ultimately reaching the value $\log\mathcal{B}_{28} = 0.0
\pm 0.12$, which indicates no preference for $N=8$ over the $N=2$
model, demonstrating that the ability of the $N=8$ model to (over)fit
the data is offset by the penalty of its increased complexity.

As mentioned earlier, we will adopt here the straightforward approach
of determining the constraints on parameters by conditioning on the
most favoured model $\hat{N}=3$, rather than performing model
averaging according to their Bayes factors. Since it is of interest to
understand any biases or constraints on the parameters imposed by our
choice of priors, we first consider the `posterior'
$\Pr({\mathbf\Theta}_{\rm c}|{\mathbf D}={\mathbf 0},\mathcal{M}_3)$
on the cluster parameters obtained in the absence of any data. This is
calculated simply by setting the likelihood to a constant value, so
that the sampler explores just the prior $\pi({\mathbf\Theta}_{\rm
  c})$. The resulting 1-D and 2-D marginalised distributions for the
sampling parameters ${\mathbf\Theta}_{\rm c}$ are shown in
figure~\ref{fig:post1} (left panel); in addition we plot the derived
parameter $Y_{\rm tot}$ defined in equation~(\ref{eq:Ycylsph}). These plots
show that we correctly recover the assumed prior distributions, and
also reveal the constraints that our choice of priors has placed on
the derived parameter $Y_{\rm tot}$.  The plots are produced using the
open source Python library {\tt corner.py} \citep{2016JOSS.00024}.
\begin{figure}
\begin{center}
\includegraphics[height=6.5cm]{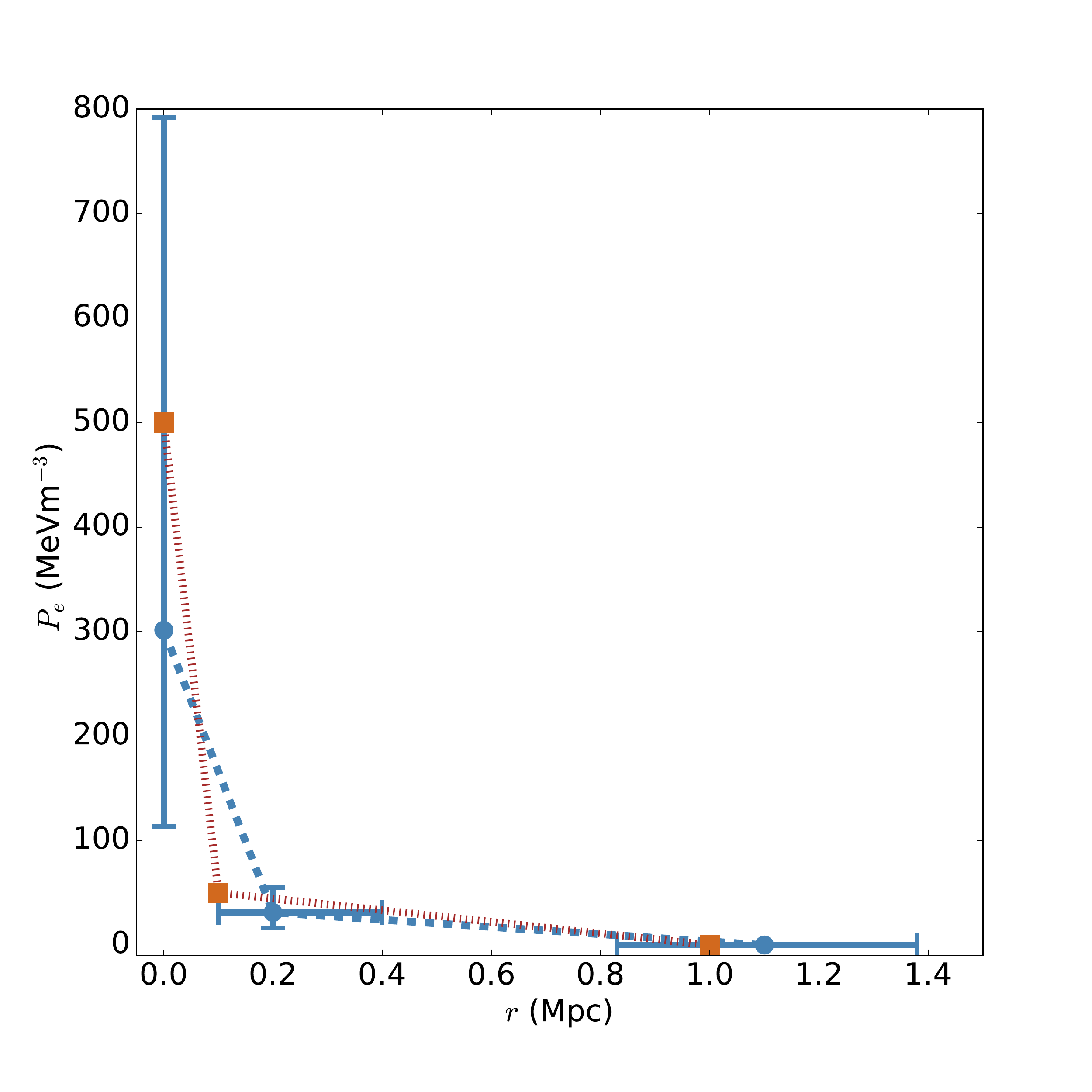}
\caption{The reconstructed pressure profile $P_{\rm e}(r)$ (dotted line),
  conditioned on $N=3$ nodes, obtained from analysis of simulation
  SIM$_1$. The position and amplitude of each node correspond to the
  mean of the corresponding marginal distribution, and the errors bars
  denote the $68\%$ Bayesian credible intervals in each
  direction; the nodes are then linearly interpolated to obtain
  $P_{\rm e}(r)$. The input pressure profile used to generate the
  simulation is also plotted (solid line).\label{fig:profile1}}
\end{center}
\end{figure}

The corresponding plot obtained after analysing the simulation SIM$_1$
is shown in figure~\ref{fig:post1} (right panel), and shows the effect
that the data have in updating the prior, via the likelihood function,
to produce the posterior, in the spirit of Bayes' theorem.  In
particular, one sees that the cluster position ($x$ and $y$) on the
sky is firmly constrained and the true values lie within few
arcseconds of the means of the posterior probability distributions for
$x$ and $y$. Turning to the parameters defining the nodal model of the
pressure profile, one sees that the amplitude $P_0$ of the first node
is not well constrained, and we are essentially recovering just the
prior distribution. This is to be expected, since an interferometric
SZ observation is insensitive to lengths scales on the sky that
correspond to Fourier modes in the visibility plane lying well below
the shortest baseline (in units of wavelengths) of the interferometer. Consequently, for this
simulation, the observations cannot probe the cluster inner core and
thus provide no information on the pressure $P_0$ at the centre. By
contrast, the position and amplitude of the second node, ($r_1$,
$P_1$), are both constrained relative to their prior distributions,
although their 2D marginal reveals a clear degeneracy between them.
The position $r_2$ of the third (and final) node is very well
constrained. Since this node corresponds to the point at which the gas
pressure drops to zero, it is a valuable quantity for defining the
extent of the gas distribution in the cluster. Indeed, obtaining a
robust constraint on this quantity can provide insight to the
dynamical state of the cluster. Finally, we note that the important
derived parameter $Y_{\rm tot}$ is also very well constrained. From
their 2-D marginal, however, one sees that there is some degeneracy
between the parameters $r_2$ and $Y_{\rm tot}$.

Rather than viewing the posterior constraints on the individual
parameters $r_1$, $r_2$, $P_0$ and $P_1$, as in figure~\ref{fig:post1}, it can
be more intuitive and instructive to plot the corresponding inference
on the reconstructed pressure profile $P_{\rm e}(r)$ directly. This
may be performed in a number of ways. For example, one may plot the
posterior probability $\Pr(P|r, {\mathbf D}, \mathcal{M}_3)$, in
normalised slices at constant $r$ \citep{2016MNRAS.455.2461H}. We will
not pursue that method here, however, and instead adopt the simpler
approach of plotting the mean position and amplitude of each node,
together with horizontal and vertical error bars representing the 68
percent marginalised Bayesian credible interval in each direction;
these nodes are then linearly interpolated to obtain the reconstructed
pressure profile, as shown in figure~\ref{fig:profile1}.
The input pressure profile used to generate the simulation is also
plotted.  As one might expect from figure~\ref{fig:post1} (right
panel), the reconstructed pressure profile is consistent with the
input profile, with the true node locations in $(r,P)$-space all lying
within the $68\%$ Bayesian credible intervals of the
corresponding inferred node locations. It is again clear, however, that
the central pressure $P_0$ is poorly constrained, as discussed above,
but that the remaining node parameters are reasonably well determined.

In the remainder of the paper, we will plot the reconstructed $P_{\rm
  e}(r)$ directly, as in figure~\ref{fig:profile1}, together with the
posterior distributions on the position ($x$ and $y$) of the cluster
and the important physical parameters $r_{N-1}\equiv r_{\rm max}$ and
$Y_{\rm tot}$.

\subsection{Simulation SIM$_2$}
\begin{figure}
\includegraphics[width=\linewidth]{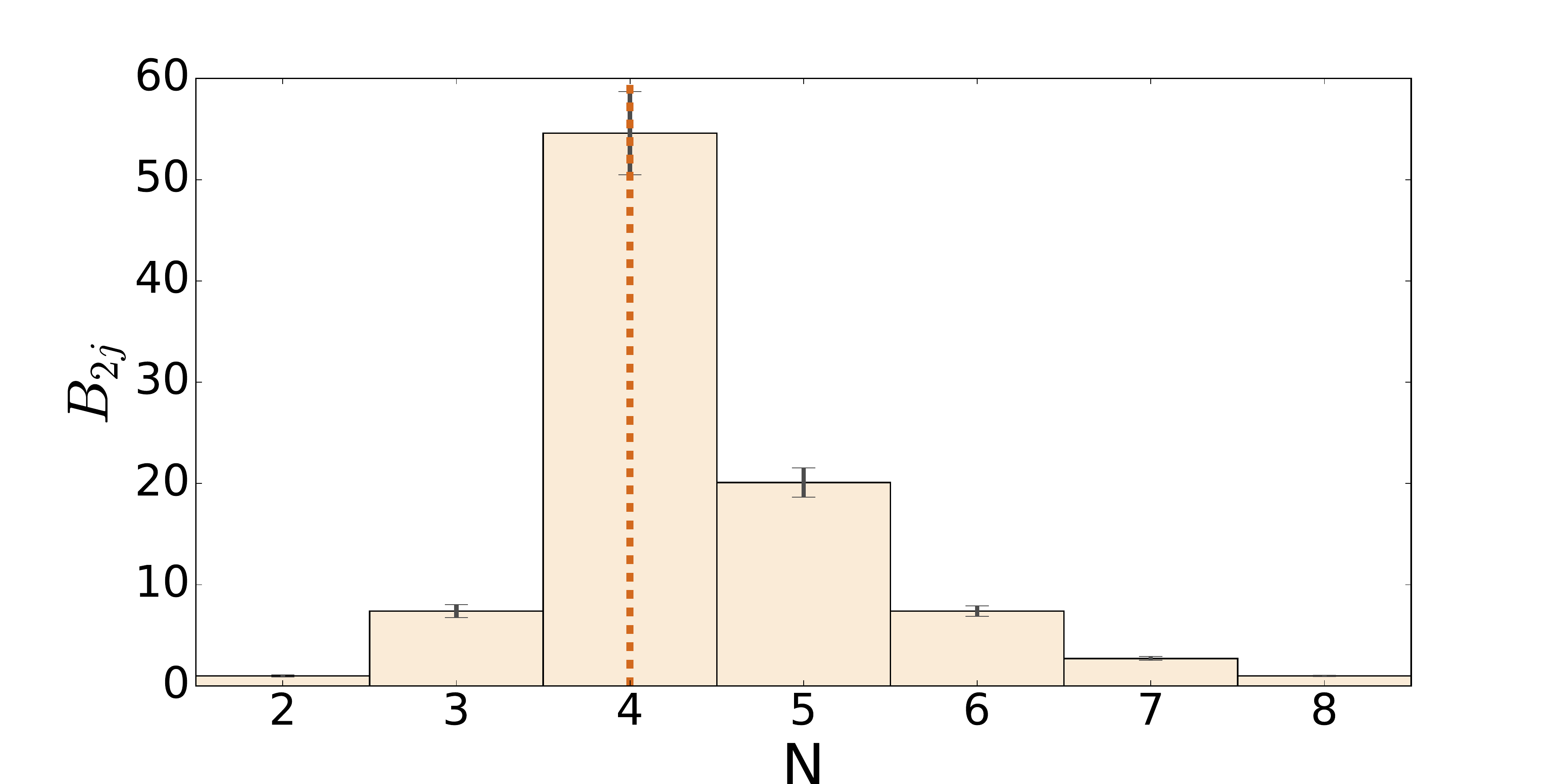}
\caption{As for figure~\ref{fig:bfhist1}, but for the analysis
of simulation SIM$_2$.\label{fig:bfhist2}}
\end{figure}
\begin{figure}
\begin{center}
\includegraphics[width=7.1cm]{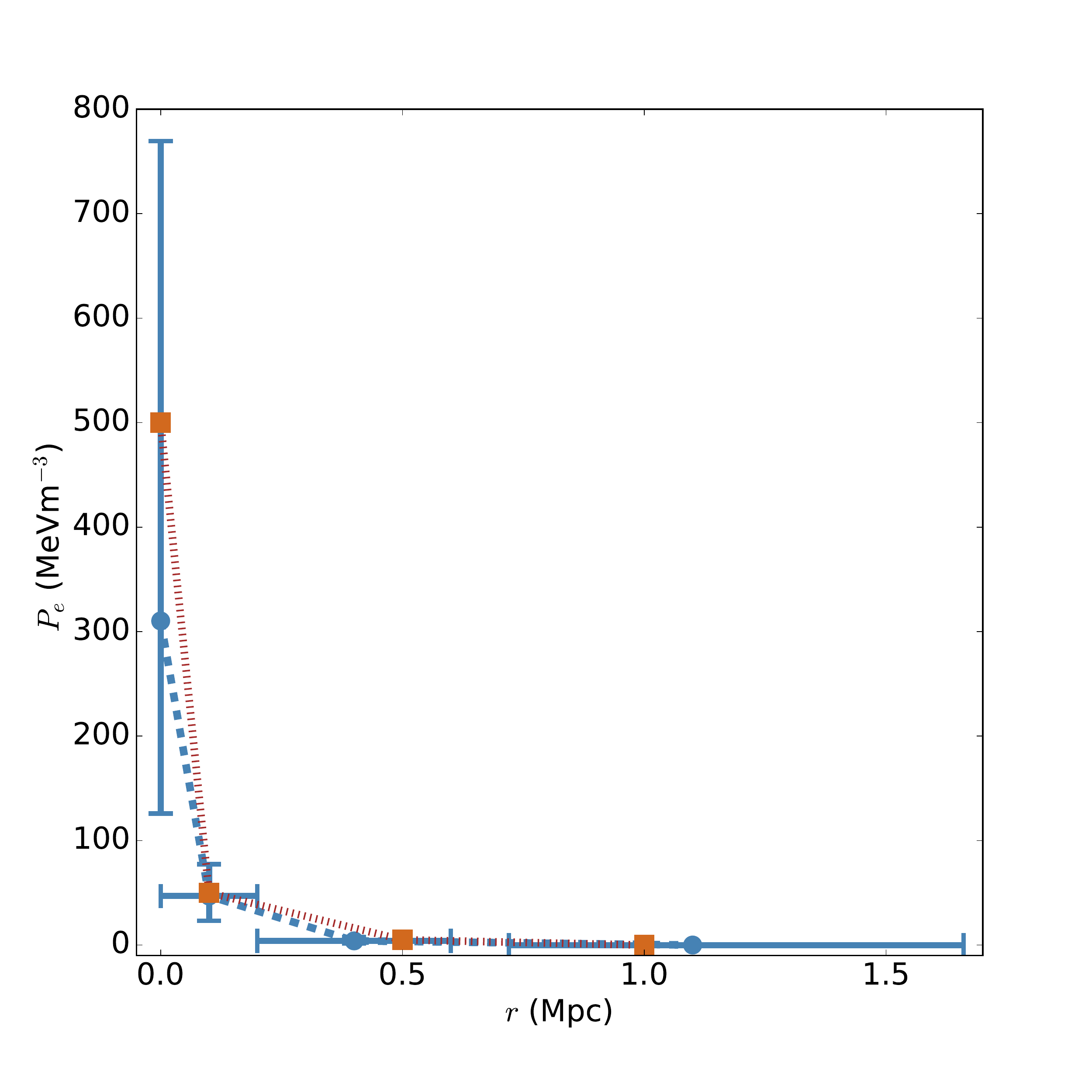}\\
\vspace{8mm}
\includegraphics[width=5.8cm]{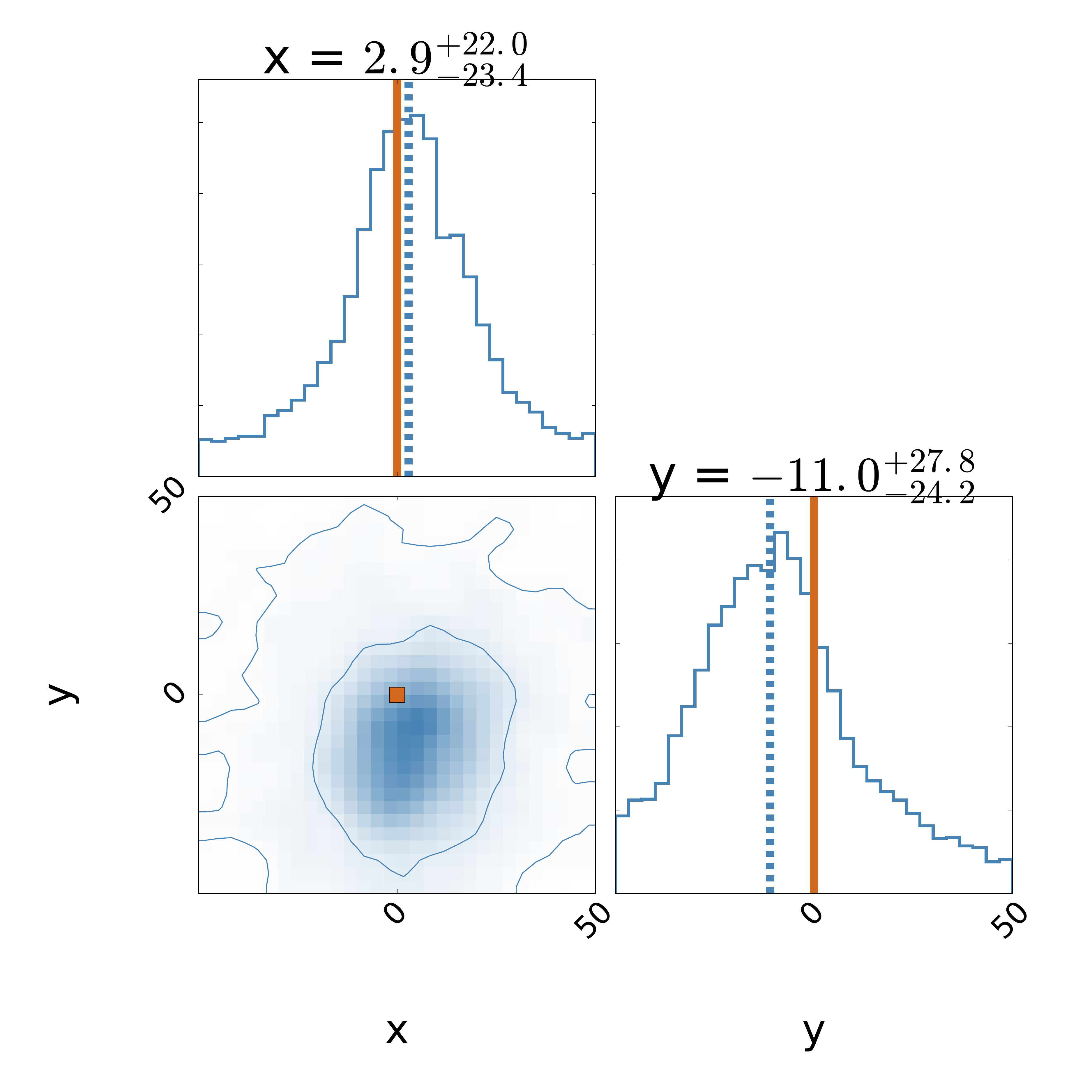}\\
\vspace{5mm}
\includegraphics[width=5.8cm]{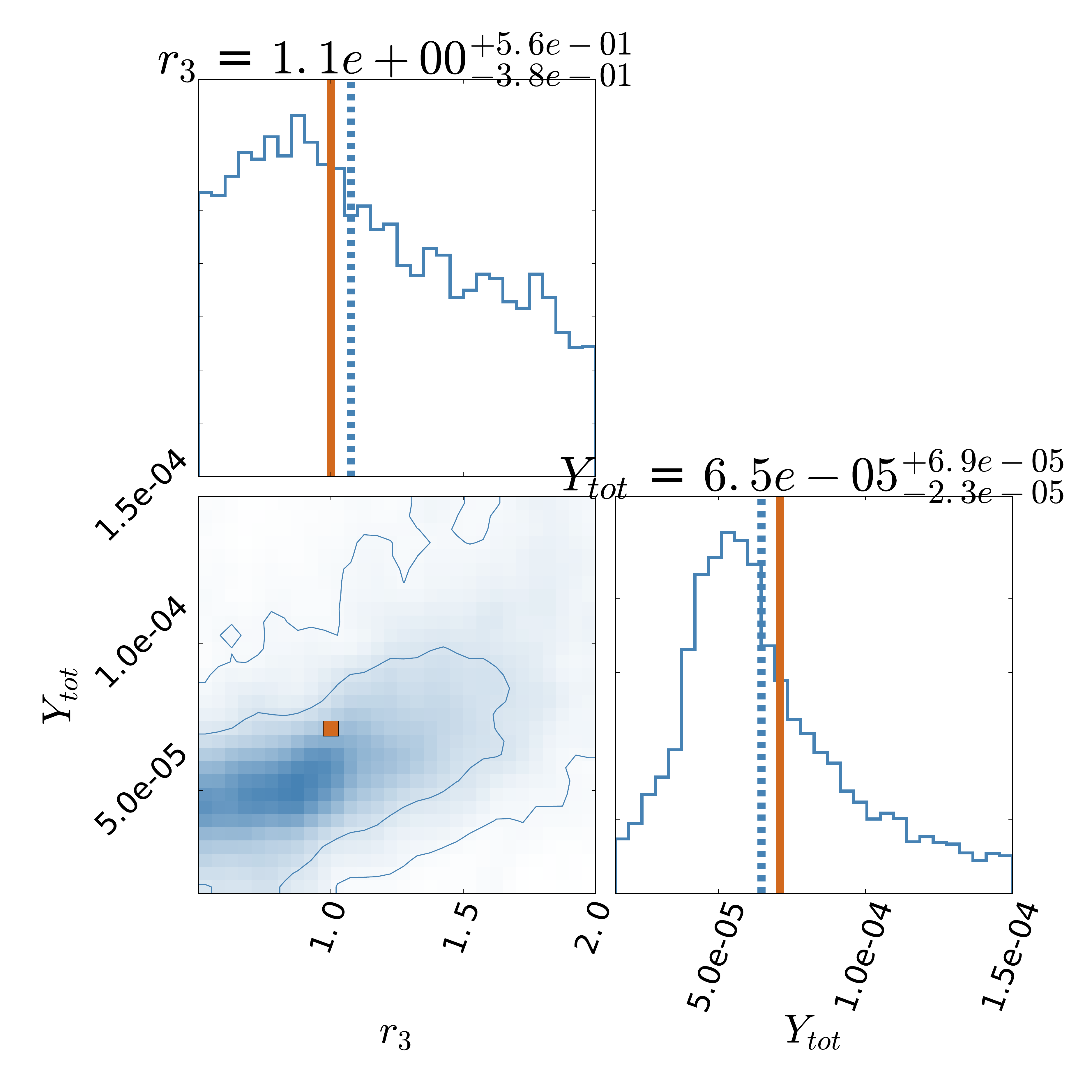}
\caption{Constraints on cluster parameters, conditioned on $N=4$
  nodes, obtained from analysis of simulation SIM$_2$. Top: the
  reconstructed pressure profile $P_{\rm e}(r)$, constructed as in
  figure~\ref{fig:post1}, together with the input profile used in the
  simulation. Middle: the cluster position
  parameters $x$ and $y$. Bottom: the gas extent $r_3$ and total
  Comptonisation parameter $Y_{\rm tot}$. 
  $x$ and $y$ are in units of arcseconds, 
$r$s are in Mpc, $P$s are in $\rm{MeVm}^{-3}$ and $Y_{\rm tot}$ is in $\rm{Mpc}^2$. 
  The contours on the 2-D
  distributions represent $68\%$ and $95\%$ Bayesian confidence
  intervals.  The vertical dotted lines show the mean values of the
  1-D distributions; these values and their $68\%$ Bayesian
  credible intervals are also quoted.  The squares and solid vertical
  lines indicate the true values of the parameters used in the
  simulation.\label{fig:post2}}
\end{center}
\end{figure}

The histogram of Bayes factors $\mathcal{B}_{2j}$ as a function of $N$
obtained in the analysis of simulation SIM$_2$ is shown in
figure~\ref{fig:bfhist2}.  Recalling that the input pressure profile for
SIM$_2$ is constructed by linear interpolation between $N=4$ nodes,
one sees that our analysis has again recovered the true value of $N$
as the most favoured. In this case, one sees the $N=2$ (straight-line)
model is again strongly disfavoured. In particular, one finds
$\ln\mathcal{B}_{24}=4.0\pm 0.12$, indicating strong evidence for the
most favoured $N=4$ model over the $N=2$ model, according to
Table~\ref{tab:bayesfactor}. The Bayes factors then gradually decline
for $N > 4$ in a similar way to that found for SIM$_1$, ultimately
reaching the value $\log\mathcal{B}_{28} = 0.0 \pm 0.12$; this again
indicates no preference for $N=8$ over the $N=2$ model, for the
reasons discussed above.

Conditioning on $N=4$ nodes, figure~\ref{fig:post2} shows the
reconstructed pressure profile $P_{\rm e}(r)$ (top panel), and the
1-D and 2-D marginal posterior distributions on the cluster position
parameters $x$ and $y$ (middle panel) and on the gas extent $r_3$ and
total Comptonisation parameter $Y_{\rm tot}$ (bottom panel).  One again
sees that the reconstructed pressure profile is consistent with the
input profile used to generate the simulation, but that cluster gas
pressure $P_0$ is poorly constrained, for the reasons we discussed
above in the context of SIM$_1$. The remaining node parameters are
again reasonably well constrained, especially the first internal node
parameters $(r_1,P_1)$.  Moreover, one again obtains tight constraints
on the cluster position, consistent with the input values. The cluster
extent $r_3$ and total Comptonisation parameter are also both well
constrained and in agreement with the input values. As in SIM$_1$,
however, the 2-D marginal distribution of $r_2$ and $Y_{\rm tot}$
reveals a mild degeneracy.

\subsection{Simulation SIM$_3$}
\begin{figure}
\includegraphics[width=\linewidth]{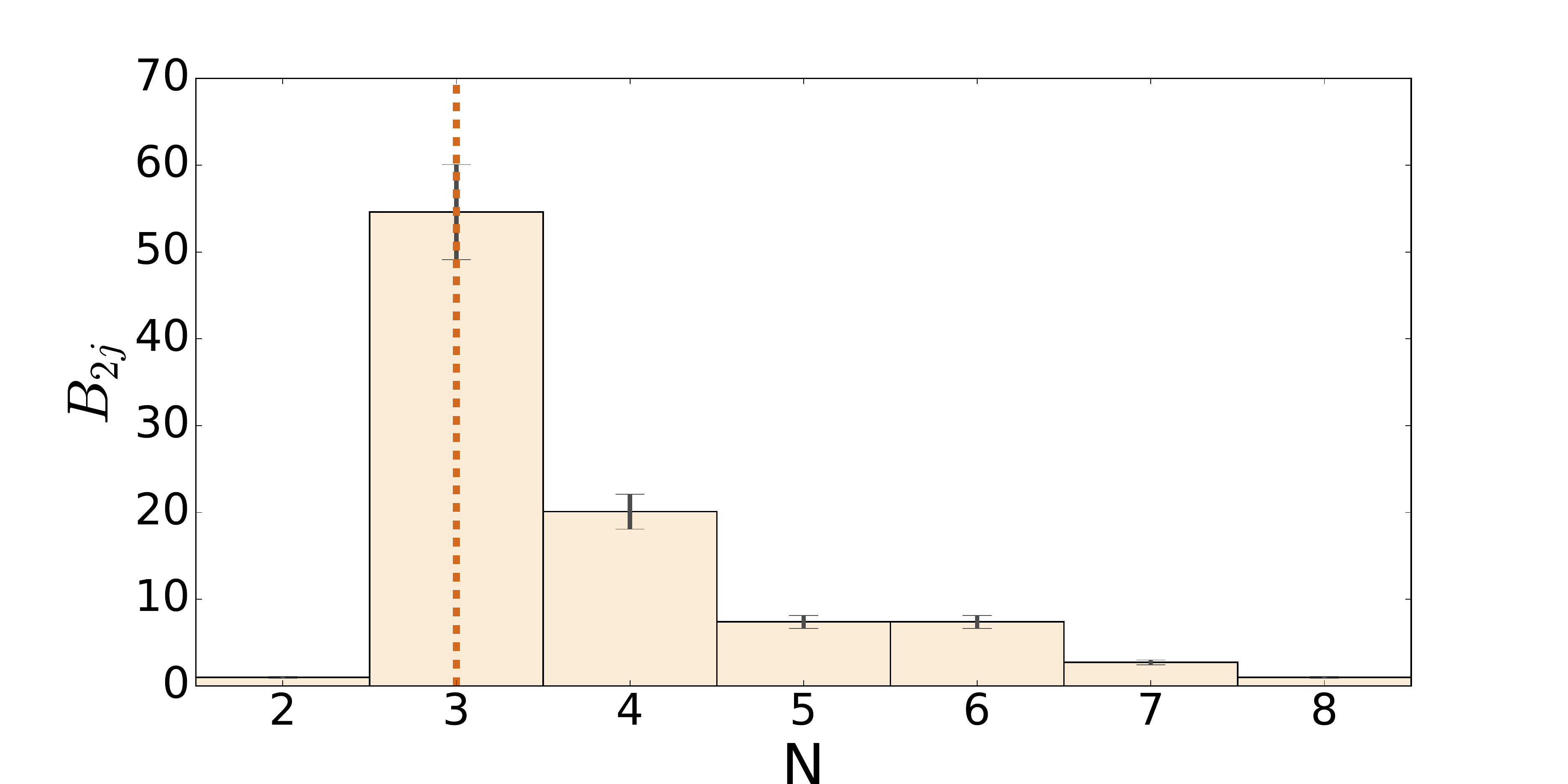}
\caption{As for figure~\ref{fig:bfhist1}, but for the analysis
of simulation SIM$_3$.\label{fig:bfhist3}}
\end{figure}
\begin{figure}
\begin{center}
\includegraphics[width=7.1cm]{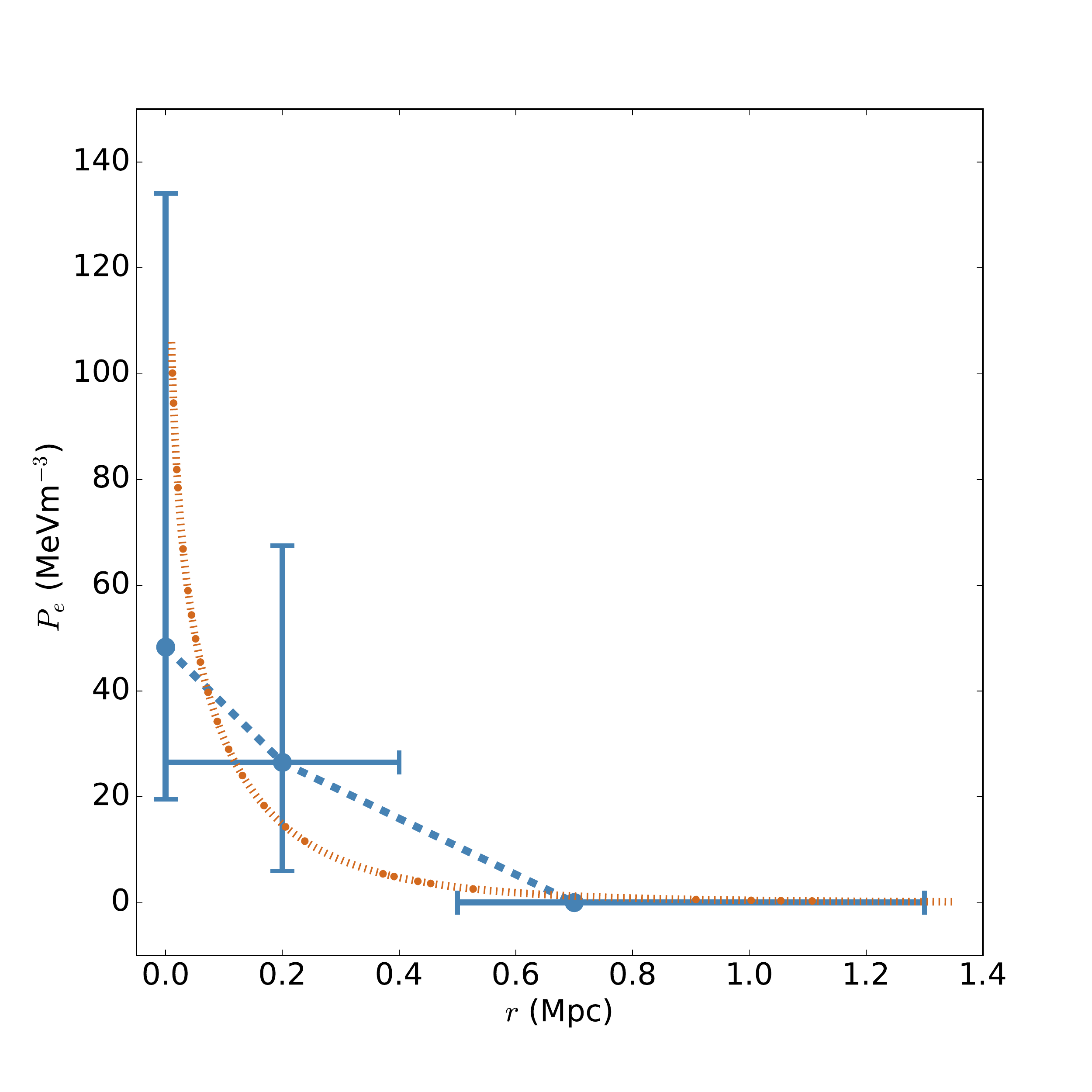}\\
\vspace{6mm}
\includegraphics[width=5.8cm]{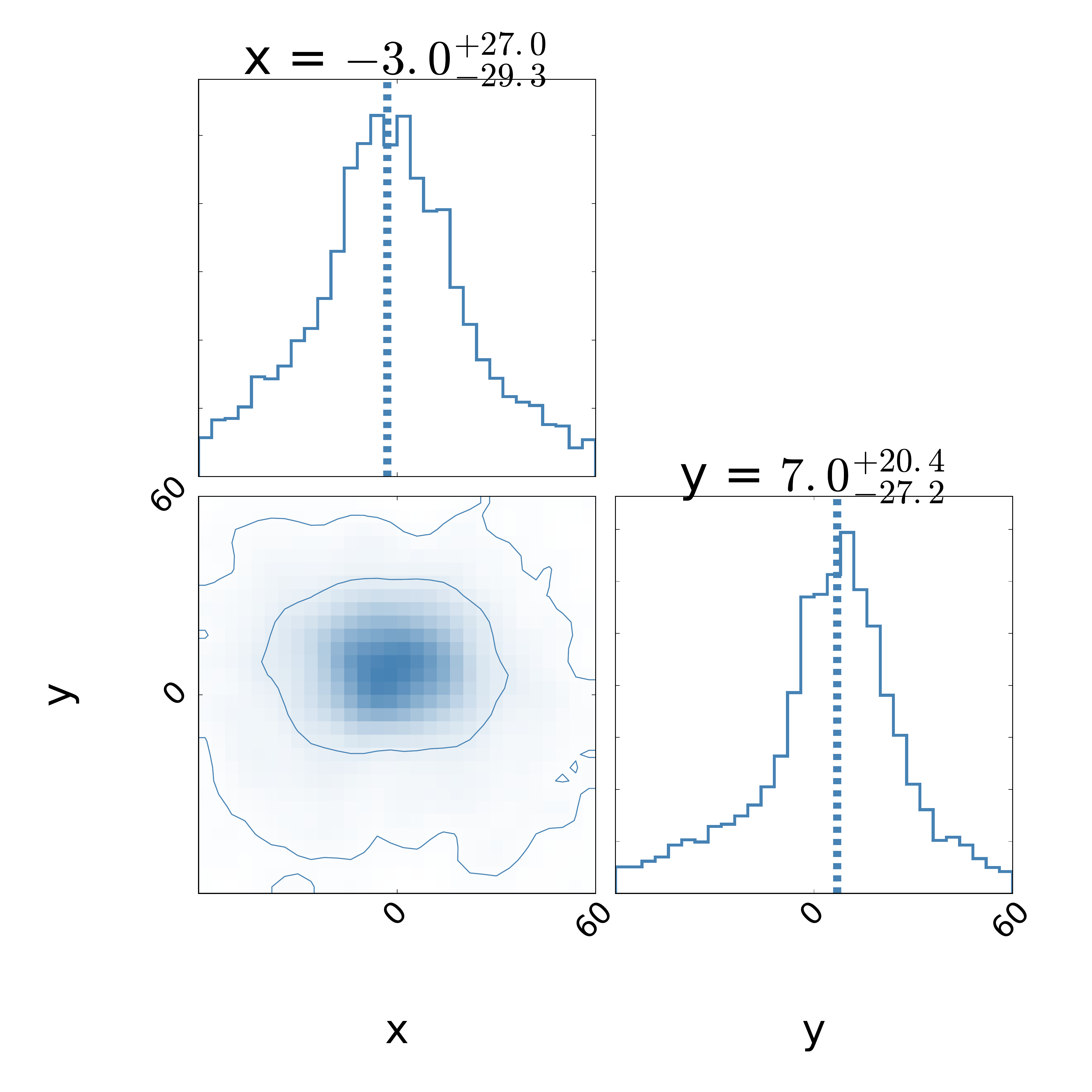}\\
\vspace{2mm}
\includegraphics[width=5.8cm]{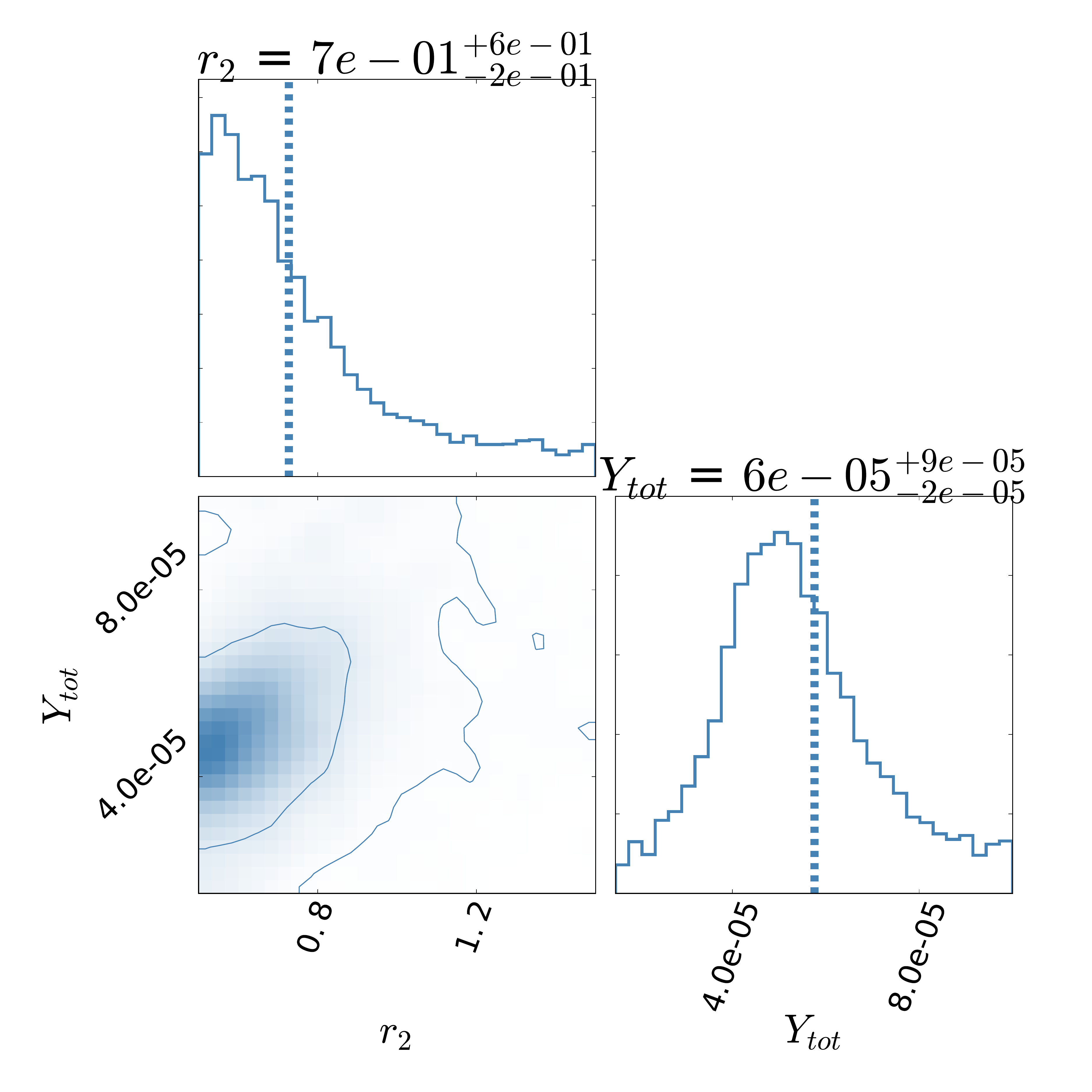}
\caption{As for figure~\ref{fig:post2}, but conditioned on $N=3$ nodes,
  and obtained from the analysis of simulation SIM$_3$.  
\label{fig:post3}}
\end{center}
\end{figure}

The histogram of Bayes factors $\mathcal{B}_{2j}$ as a function of $N$
obtained in the analysis of simulation SIM$_3$ is shown in
figure~\ref{fig:bfhist3}.  The input pressure profile for SIM$_2$ is not
constructed from a linear interpolation between nodes, but instead
from the cluster model of \cite{2012MNRAS.423.1534O,
  2013MNRAS.430.1344O}, which assumes that the pressure is described
by the generalised NFW (GNFW) profile
\citep{2007ApJ...668....1N}. Hence, in this case, there is no
`correct' number of nodes to recover. Instead, the most favoured value
of $\hat{N}=3$ nodes gives an indication of the level of complexity in
the pressure profile reconstruction that is supported by the data.
Thus, for interferometric SZ observations of the type simulated, the
data support only a very simple reconstruction of the pressure
profile, favouring a representation consisting of just two
straight-line segments. Indeed, the variation of the Bayes factor with
$N$ is very similar to that shown in figure~\ref{fig:bfhist1} for SIM$_1$, for
which the input pressure profile had precisely this simple form.

Conditioning on $N=3$ nodes, figure~\ref{fig:post3} shows the
reconstructed pressure profile $P_{\rm e}(r)$ (top panel), and the
1-D and 2-D marginal posterior distributions on the cluster position
parameters $x$ and $y$ (middle), and the gas extent $r_3$ and total
Comptonisation parameter $Y_{\rm tot}$ (bottom). Although in this
simulation there are no `correct' locations for the nodes in
$(r,P)$-space, one sees that the reconstructed node locations yield
reconstructed pressure profiles that are consistent with the input
one, although the plotted `best-fit' profile is somewhat shallower
than the GNFW profile assumed in the simulation. One can understand
this behaviour by recalling that the SZ Comptonisation parameter given
in (\ref{eq:Ycylsph}) is proportional to the integral of $P_{\rm e}(r)$ along
the line-of-sight through the cluster. Thus, the larger amplitude of
the first (only) internal node relative to the input profile can
offset the smaller amplitude of the node at the origin. Once again our
analysis serves to highlight the rather coarse level of detail in the
reconstructed pressure profile that is achievable with SZ observations
of the type simulated. Nonetheless, one still obtains tight
constraints on the cluster position, which are consistent with the
input values, and also on the cluster extent $r_3$ and total
Comptonisation parameter. We again see, however, that there is a
slight degeneracy in the 2-D marginal distribution of $r_2$ and
$Y_{\rm tot}$.

\subsection{Cluster MACS$\,$J0744+3927}
\begin{figure}
\includegraphics[width=\linewidth]{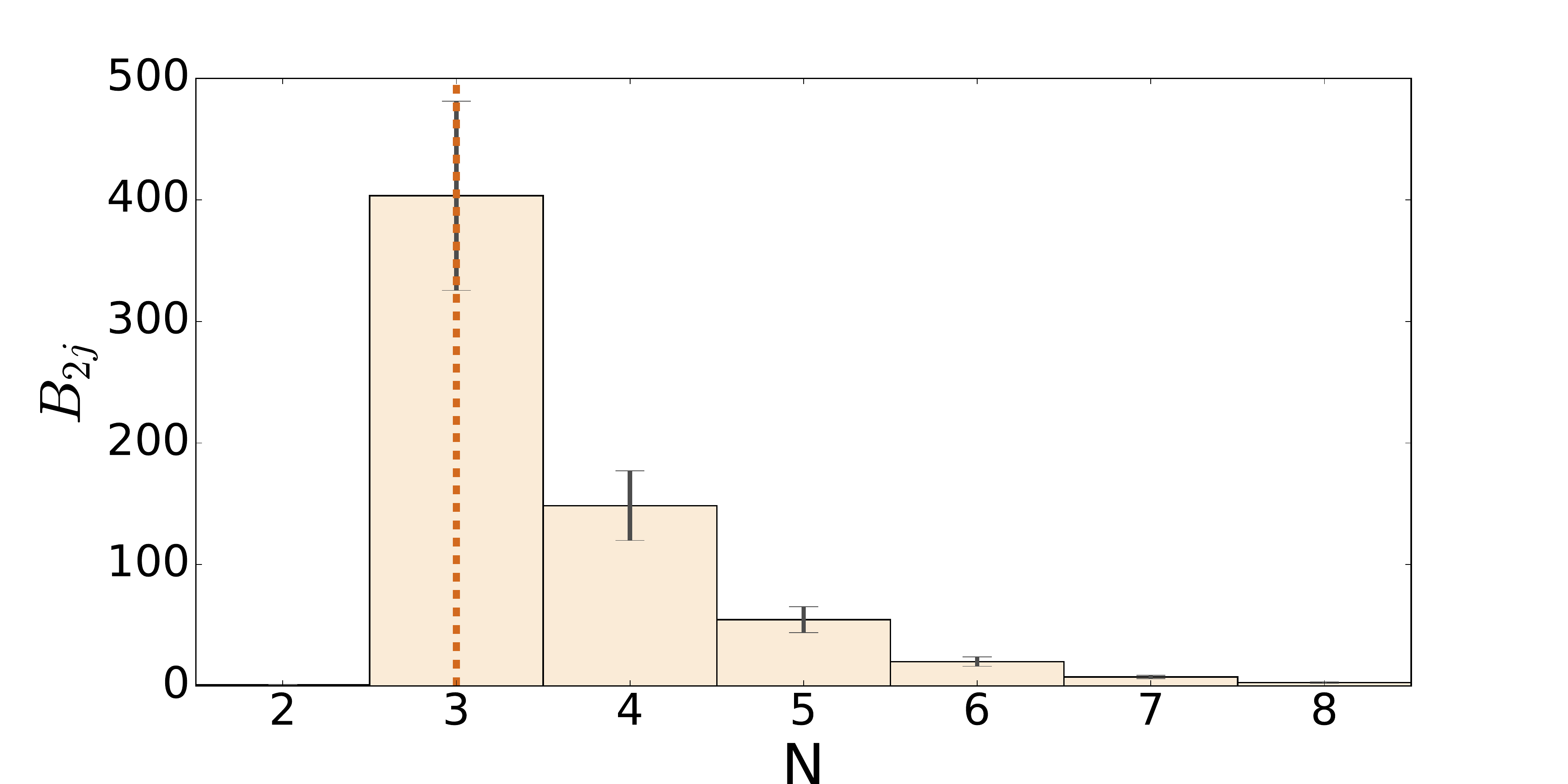}
\caption{As for figure~\ref{fig:bfhist1}, but for the analysis
of real AMI observations of the cluster MACS$\,$J0744+3927.\label{fig:bfhistMAJ}}
\end{figure}
\begin{figure}
\begin{center}
\includegraphics[width=7cm]{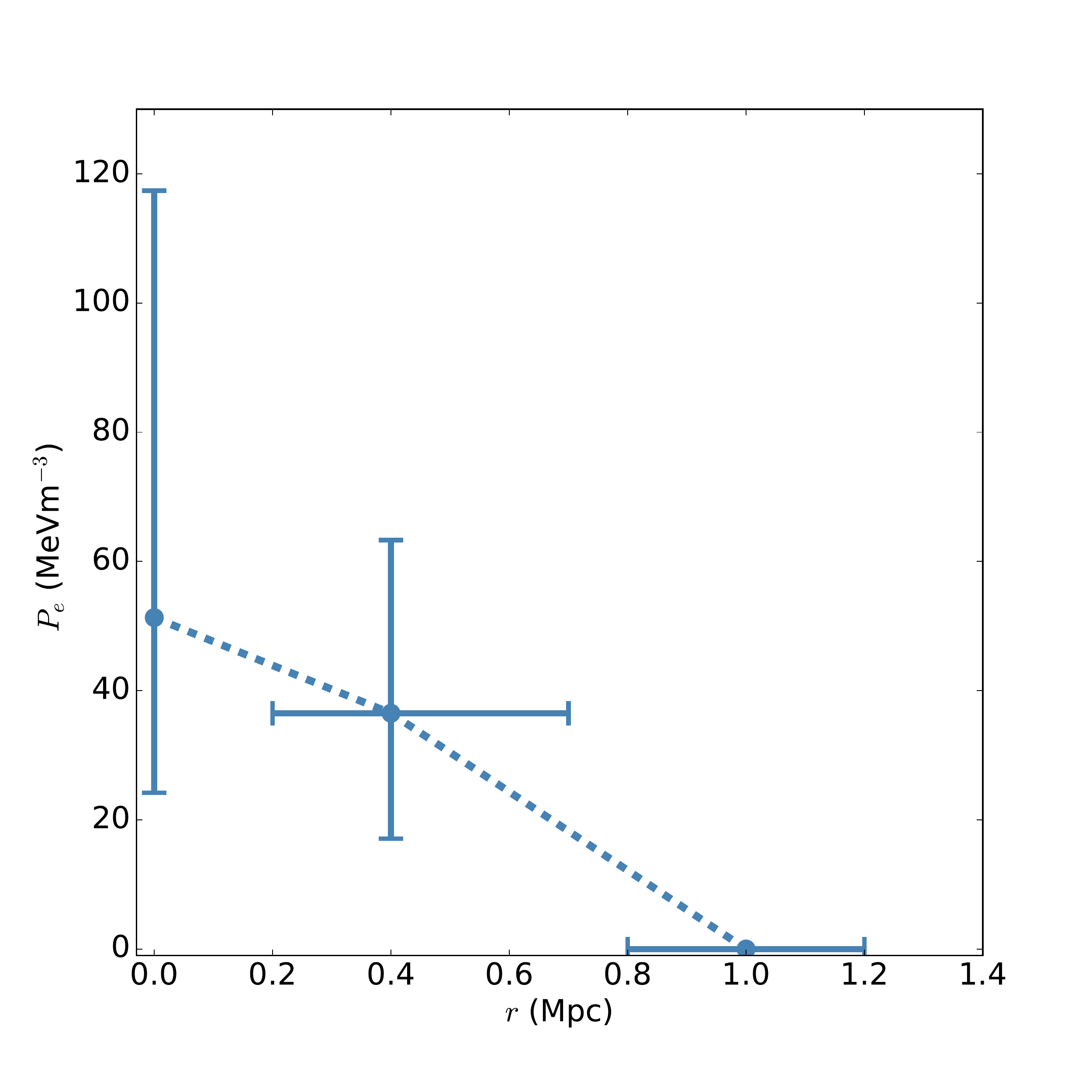}\\
\vspace{6mm}
\includegraphics[width=5.7cm]{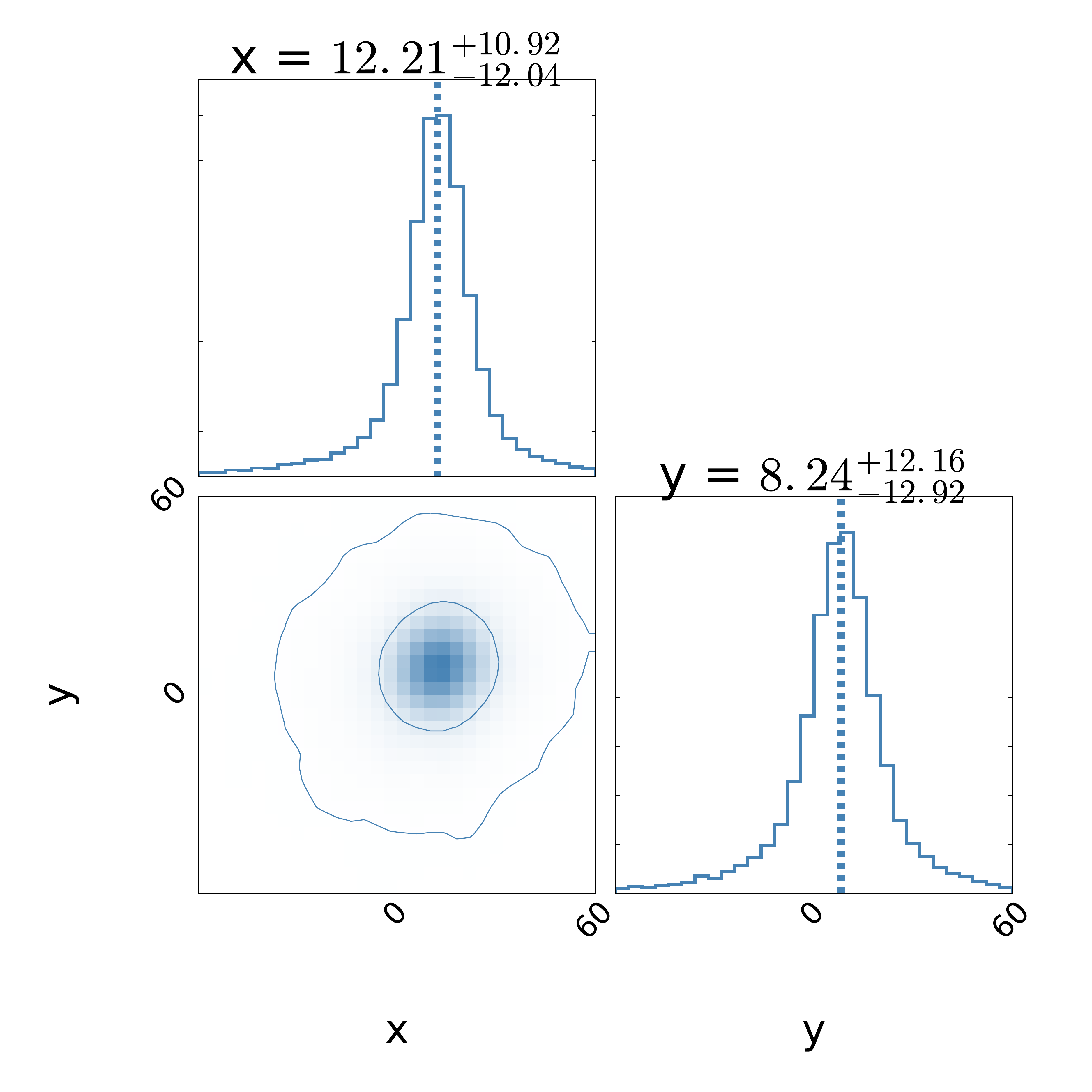}\\
\vspace{2mm}
\includegraphics[width=5.7cm]{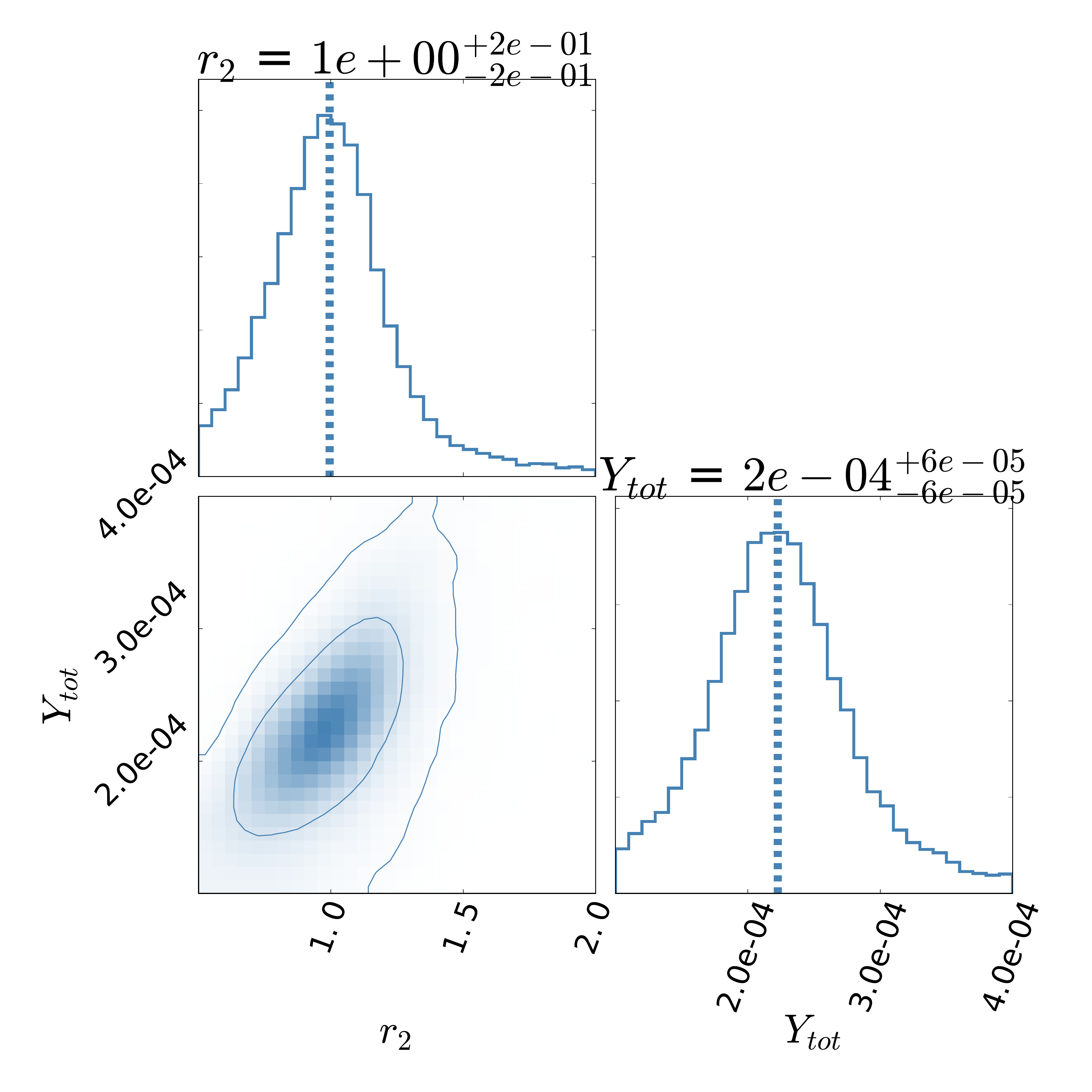}
\caption{As for figure~\ref{fig:post3}, but conditioned on $N=3$ nodes,
and obtained from the analysis of real AMI observations of the cluster
MACS$\,$J0744+3927.\label{fig:postMAJ}}
\end{center}
\end{figure}

The histogram of Bayes factors $\mathcal{B}_{2j}$ as a function of $N$
obtained in the analysis of real AMI observations of the cluster
MACS$\,$J0744+3927 is shown in figure~\ref{fig:bfhistMAJ}.  For this real
cluster, one sees that the variation of the Bayes factor
$\mathcal{B}_{2j}$ is broadly similar to that shown in
figure~\ref{fig:bfhist3}, obtained from the analysis of simulation
SIM$_3$, for which an input GNFW pressure profile was assumed. In
particular, the most favoured model again has $N=3$ nodes, after which
the Bayes factors gradually decline with increasing $N$.  This
indicates that the data support a model for the pressure profile that
is no more complex than two straight-line segments. It is worth
noting, however, that in comparing the favoured $N=3$ model with the
base $N=2$ (straight-line) model, one obtains $\ln\mathcal{B}_{23} =
6.0 \pm 0.2$, which corresponds to a decisive favouring of the former
model, according to Table~\ref{tab:bayesfactor}. Thus, one may deduce
at high confidence from our analysis, in a model independent manner,
that the pressure profile is not simply a linear function of $r$.

Conditioned on $N=3$ nodes, figure~\ref{fig:postMAJ} shows the
reconstructed pressure profile $P_{\rm e}(r)$ (top panel), and the 1-D
and 2-D marginal posterior distributions on the cluster position
parameters $x$ and $y$ (middle), and the gas extent $r_2$ and total
Comptonisation parameter $Y_{\rm tot}$ (bottom).  Once again the
central pressure $P_0$ is poorly constrained, but the remaining node
locations in $(r,P)$-space are better determined. It is interesting
that the resulting reconstructed pressure profile is 
convex, rather than concave such as those obtained from the simulated
observations, although the uncertainty in the location of the single
internal node also allows for concave profiles. It is also worth
recalling from our analysis of simulated data how SZ observations of
this type allow for only a very coarse reconstruction of the cluster
pressure profile, owing to the SZ effect being proportional to the
line-of-sight integral of the pressure. Indeed, from
figure~\ref{fig:post3}, we recall that in the analysis of simulation
SIM$_3$ the reconstructed pressure profile was somewhat flatter than
the input GNFW profile, although still consistent with it to within
the error-bars, and a similar effect could be occurring in
figure~\ref{fig:postMAJ}.

Nonetheless, as we found in our analysis of simulated data, the
cluster position ($x$ and $y$) is well constrained, and the
corresponding 2-D marginal posterior distribution shows no sign of 
degeneracy. The gas extent
$r_3$ and total Comptonisation parameter $Y_{\rm tot}$ are also both
very well determined, although their 2-D marginal posterior distribution shows that same
slight degeneracy as seen in the analyses of the simulation
observations.

\section{Conclusions}
\label{sec:concs}

Almost all current approaches to modelling observations of galaxy
clusters rely on assuming some parameterised functional form for the
properties of the cluster, such as gas density, dark matter density or
temperature. A generic weakness of this approach is that these
functional forms have usually been arrived at through empirical means,
via the analysis of $N$-body simulations or observations, and are
often chosen to have simple analytic expressions, rather than being
fundamental or physically well motivated.

In this paper, we have moved away from this approach and presented a
free-form model for the physical properties of galaxy
clusters. Previous attempts to model clusters in this way have
typically relied simply on dividing the cluster into a predefined
number of cells, or concentric shells for spherical clusters, and
determining the value of each physical quantity of interest within
these subregions  \citep{2015A&A...574A.122T}. Such approaches typically lead to
under-determined inverse problems that therefore need to be
regularised in some way.  There is considerable freedom in how to
choose the level or nature of the regularisation to apply, and the
results can vary significantly depending on how this choice is
made. We have therefore presented an alternative approach to free-form
reconstruction in which the complexity of the model is determined
directly from the data.  This is achieved by representing each
independent cluster property as some interpolating or approximating
function that is specified by a set of control points, or `nodes', for
which the number of nodes, together with their positions and
amplitudes, are allowed to vary and are inferred from the data in a
Bayesian manner, employing both model selection and parameter
estimation.

To demonstrate our approach in a simple setting, we have applied it to
the particular case of modelling interferometric SZ observations of
spherical galaxy clusters. In this context, the free-form part of
the cluster model is simply a nodal representation of the electron
pressure profile $P_{\rm e}(r)$. We have performed Bayesian analyses
of simulated observations with the Arcminute Microkelvin Imager (AMI)
of three separate model clusters. 

In the first two simulations, the input pressure profile has the same
form as that assumed in the analysis, namely a linear interpolation
between a set of $N$ nodes (with $N=3$ and $N=4$, respectively). We
showed that, in both cases, our Bayesian model selection analysis
returned the true value of $N$ as the most favoured. Moreover the
resulting reconstructed pressure profiles were consistent with those
used as input. In our third simulation, in which the input pressure
was assumed to follow a GNFW, the most favoured model again had $N=3$
nodes, and the resulting reconstructed pressure profile was consistent
with the input one.  In all cases, we found that the central pressure
of the cluster is not well determined, since interferometric
observations of the type simulated do not probe length scales
corresponding to the inner core.  In the analysis of our third
simulation, we also noted that the reconstructed pressure profile was
somewhat shallower that the singular GNFW profile used to generate the
simulation (although still consistent with it), which results from the
SZ effect being proportional only to the line-of-sight integral of the
pressure in the cluster. A general feature of our results is that SZ
interferometric observations of this type allow for only a very coarse
reconstruction of the cluster pressure profile. Nonetheless, we also
find that in all cases one obtains tight constraints on the cluster
position, and that the cluster extent and total Comptonisation
parameter are also both well determined.

We also applied our approach to real AMI observations of the cluster
MACS$\,$J0744+3927. We found that the most favoured model has $N=3$ nodes,
and that the corresponding best-fit reconstructed pressure profile is
convex, although the uncertainties of the node locations in
$(r,P)$-space also allow for concave profiles. As we found in the
analysis of simulations, the central pressure is poorly determined but
the remaining node parameters are reasonably well constrained. Once
again, we found that cluster position, cluster gas extent and total
Comptonisation parameter are all very well constrained.

In closing, some further general points and avenues for future
research are worth discussing. First, the tight constraints obtained
on the cluster position and on the two very important cluster
parameters $r_{\rm max}$ and $Y_{\rm tot}$ demonstrate the robustness
of our approach. Moreover, with only minor modification, the method
may prove very useful in cluster detection. Although, for the sake of
illustration, we assumed the cluster redshift in our analyses
presented here, this is not necessary. One can easily re-perform the
analysis by instead constructing a nodal model for the pressure
profile $P_{\rm e}(\theta)$, where $\theta$ is the projected angle on
the sky from the centre of the cluster. In this way, the approach does
not depend on the redshift, but will still produce the tight
constraints on the cluster position, angular extent and $Y_{\rm tot}$.
Finally, since our Bayesian approach to the inference produces
posterior weighted samples in the parameter space, further directions
for future development include defining other derived parameters that
capture particular features of interest in the pressure profile. One
example, motivated by our analysis of the cluster MACS$\,$J0744+3927, would
be a statistic that embodies the concavity or convexity of the
pressure profile. Others might include a parameter that quantifies
the cuspy versus core nature of the central region of the cluster.  In any
case, one may easily use the posterior samples to determine the full
(joint) posterior distribution of such derived parameters.

\section*{Acknowledgments}
The authors thank William Handley for illuminating
discussions regarding Bayesian inference.  This work was
performed using both the Darwin Supercomputer of the University of
Cambridge High Performance Computing Service
(http://www.hpc.cam.ac.uk/), and COSMOS Shared Memory system at DAMTP,
University of Cambridge operated on behalf of the STFC DiRAC HPC
Facility. Darwin Supercomputer is provided by Dell Inc. using
Strategic Research Infrastructure Funding from the Higher Education
Funding Council for England and funding from the Science and
Technology Facilities Council.  COSMOS Shared Memory system is funded
by BIS National E-infrastructure capital grant ST/J005673/1 and STFC
grants ST/H008586/1, ST/K00333X/1. We are grateful to Stuart Rankin
and COSMOS management team for their computing assistance. MO thanks
Astro Hack Week 2016 for valuable discussions and insights on Bayesian
inference and statistics. YCP acknowledges support from a 
Trinity College Junior Research Fellowship.
\setlength{\labelwidth}{0pt} 

\label{lastpage}
\end{document}